\documentclass[twocolumn]{aastex63}

\usepackage{natbib}
\usepackage{graphicx}
\usepackage{epstopdf}

\usepackage{amssymb}
\usepackage{amsmath}
\usepackage{ulem}
\usepackage{enumitem}
\usepackage{float}

\usepackage{mathrsfs} 
\usepackage{txfonts}

\newcommand{\cha}[1]{#1}
\newcommand{\bb}[1]{\textbf{#1}}

\newcommand{\be}{\begin{equation}}
\newcommand{\ee}{\end{equation}}
\defcitealias{MGV18}{MGV18}
\defcitealias{VG16}{VG16}
\received{xx}
\revised{xx}
\accepted{xx}

\begin{document}

\title{Comparing turbulent cascades and heating vs spectral anisotropy in solar wind via direct simulations
\footnote{Released on }}

\correspondingauthor{Victor Montagud-Camps}
\email{Victor@aurora.troja.mff.cuni.cz}

\author{Victor Montagud-Camps}
\affiliation{Charles University, Faculty of Mathematics and Physics, V Holesovickach 2, 180 00 Prague 8, Czech Republic}

\author{Roland Grappin}
\affiliation{LPP, Ecole Polytechnique, CNRS, France}

\author{Andrea Verdini}
\affiliation{Universit\`a di Firenze, Dipartimento di Fisica e Astronomia, Firenze, Italy}
\affiliation{INAF, OAA, Firenze, Italy}




\begin{abstract}
In a previous work (MGV18), we showed numerically that the turbulent cascade generated by quasi-2D structures (with wave vectors \cha{mostly-perpendicular} to the mean magnetic field) is able to generate a temperature profile close to the one observed in solar wind ($\simeq 1/R$) in the range 0.2 $\le R \le$ 1~au.
Theory, observations and numerical simulations point to another robust structure, the radial-slab, with dominant wave vectors along the radial: we study here the efficiency of the radial-slab cascade in building the $1/R$ temperature profile.
As in MGV18, we solve the \cha{three-dimensional} MHD equations including expansion to simulate the turbulent evolution. 
We find that an isotropic distribution of wave vectors with large cross helicity at 0.2 au, along with 
a large wind expansion rate, lead again to a temperature decay rate close to $1/R$ but with a radial-slab anisotropy at 1 au.
Surprisingly, \cha{the turbulent cascade concentrates in the plane transverse to the radial direction, displaying 1D spectra with scalings close to $k^{-5/3}$ in this plane}.
This supports both the idea of turbulent heating of the solar wind, and the existence of two different turbulent cascades, quasi-2D and radial slab, at the origin of the heating.

We conclude that sampling the radial spectrum in the solar wind may give but a poor information on the real cascade regime and rate when the radial slab is a non-negligible part of turbulence.
\end{abstract}



\section{Introduction}
The solar wind is a spherically expanding and turbulent flow in which the proton temperature decreases slower than the adiabatic prediction $T\propto R^{-3/4}$, where $R$ is the heliocentric distance beyond 0.3~AU. Measurements of the temperature profile return different scaling laws, $T\propto R^{-\alpha}$, with $\alpha\in[0.5,1]$ depending on the periods analyzed (at minimum or maximum of the solar cycle) or on the type of stream considered (fast, slow, or fast and slow streams) or on the distinction between parallel and perpendicular temperatures \citep{1982JGR....87...52M,1995JGR...100...13T,2011JGRA..116.9105H,2012JGRA..117.9102E,2013JGRA..118.1351H}. Despite their differences, the scaling-law exponents indicate that heating must be supplied to protons in their way out from the Sun. Interestingly, when streams are separated according to their speed and rectified for their acceleration in the heliosphere, their temperature decreases as $1/R^{0.9\pm0.1}$ (\citealt{1995JGR...100...13T, 2020ApJS..246...62M}, and specifically for fast winds \citealt{2019MNRAS.483.3730P}),
suggesting that a similar strong heating is at work in fast and slow streams. 

A possible source of slow proton cooling is a turbulent cascade, which is suggested by the power-law energy spectra currently observed since \cite{1968ApJ...153..371C}. 
In a recent work \citep[][MGV18 hereafter]{MGV18}, we obtained by direct numerical simulations a 1/R temperature decay in the distance range [0.2, 1] au, close to the slow temperature decrease of \cite{1995JGR...100...13T}. 
The numerical demonstration relies on choosing initial parameters such as the relative expansion rate, the turbulent Mach number and, most importantly, \cha{the initial eddy geometry that characterizes the strong cascade, namely a quasi-2D geometry}.

\cha{Solar wind turbulence is not made only of the quasi-2D geometry, but an important contribution is often given by the so-called slab component 
\citep{1990JGR....9520673M,1992JPlPh..48...85Z,2005ApJ...635L.181D,2017ApJ...835..147Z}. 
In the slab component the wave vectors are mostly along the mean magnetic field $B_0$, while they lay mostly within the plane perpendicular to $B_0$ in the quasi-2D component\footnote{With quasi-2D component, or geometry, we refer to only wavevector anisotropy, with no further specification on the component (or variance) anisotropy}.}

\cite{2005ApJ...635L.181D} found that at a 20 minutes scale, the 2D component is dominant in the slow wind, while the slab is dominant in the fast wind. 
In line with this work, we supposed in \citetalias{MGV18} that the turbulent heating obtained in the our simulations of the \cha{quasi-2D} regime applied in slow winds.
Other properties, as the spectral index and cross helicity \citep{1991AnGeo...9..416G,2013ApJ...770..125C} have been found to vary with wind speed.
Such variations should motivate investigating whether the 1/R temperature decay still obtains when passing from the quasi-2D to the slab structure.

However, there are indications that the true alternative structure in solar wind turbulence is not the slab, but the radial-slab structure.
Numerical work \citep{1993PhRvL..70.2190G},\citep[][VG16 hereafter]{VG16} and observational work at 12 hours scale \citep{Saur:1999gy} find actually that turbulence is made of a combination of \cha{quasi-2D component} and radial-slab, the latter being made of wave vectors aligned with the radial direction, as expected from the simple linear effect of wind expansion, combined with turbulence (\cite{Dong:2014fi}, \citetalias{VG16}).

Arguments against the existence of the slab structure are that there is no proof, either theoretical (but see \citealt{2017ApJ...835..147Z}) or numerical (see the negative results by \cite{Ghosh_1998_0,Ghosh_1998_1}) that the slab geometry exists in MHD turbulence, and direct observation doesn't exist, since it relies completely on the hypothesis of axial symmetry about the mean magnetic field, which is not proven.
We discuss at the end of the paper the possibility proposed by VG16 that the slab actually hides a radial-slab structure.

We thus investigate here at which conditions the observed slow 1/R temperature decay can be obtained in general by a combination of quasi-2D and radial-slab. 
We use the same method as in \citetalias{MGV18}, direct simulations of the MHD equations, including wind expansion (expanding box model or EBM, see \citealt{1993PhRvL..70.2190G,Grappin:1996ey,Dong:2014fi}). 
Within this description, visco-resistive terms dissipate kinetic and magnetic fluctuations. Our work is based on the assumption that in a direct turbulent cascade, the energy cascade rate is independent of the detailed energy dissipation mechanism. This principle is supported by recent MHD-Hall simulations (Papini et al. in preparation). 

We solve the primitive MHD equations in the heliocentric distance range $0.2 < R < 1$~au. 
We act on the heating rate by controlling three parameters, namely, the initial spectral anisotropy, the relative expansion rate (i.e., measured in terms of large eddy turnover rate), the turbulent Mach number and cross helicity.
A comparison with Helios data allows us to determine the upper scale of our simulations.

We find that the radial-slab geometry allows to generate a 1/R temperature decrease (thus close to the  measurement of \citealt{1995JGR...100...13T}) using parameters relatively close to those used for the quasi-2D turbulent geometry, but with important differences in 1D scaling laws. 

Section 2 introduces the equations, parameters and basic physics. Section 3 describes the initial conditions of the simulations. Section 4 gives the main results. Section 5 is a discussion.


\section{Equations, basic parameters and physics} 

\begin{figure}
\begin{center}
\includegraphics [width=1.\linewidth]{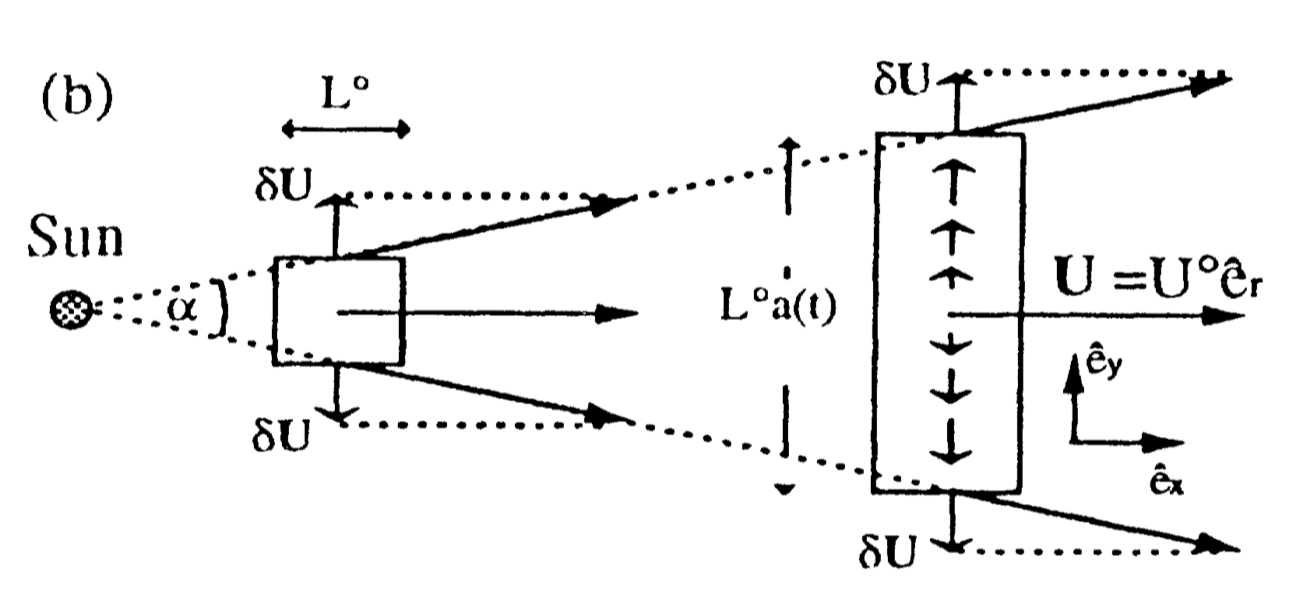}
\caption{
Scheme of the numerical domain in an expanding box model simulations. Left Box: initial state of the domain at 0.2AU; right box: final state of the domain at 1AU. The origin of the reference 
frame is at the center of the box and follows it with its transport by the wind. (From \cite{1993PhRvL..70.2190G}).
}
\label{febm}
\end{center}
\end{figure}

\subsection{Equations}
We derive here briefly the basic equations (see \citetalias{MGV18} for a detailed description, in particular, of the dissipation terms).
We start with the MHD equations for the density $\rho$, (isotropic) pressure $P$, the velocity fluctuation $u=U-U_0\hat e_r$ (where U is the total velocity and $U_0$ is the mean radial flow amplitude, and the magnetic field $B$.
Consider a Cartesian frame with X,Y,Z coordinates, the X-axis parallel to the radial passing through the middle of the box, and change to a Galilean frame moving with the mean wind along the radial coordinate.
In this frame, the plasma volume is uniformly stretched in the transverse directions (see fig.~\ref{febm}), thus neglecting curvature terms.

All fields are assumed periodic in the comobile coordinates $x,y,z$:
\begin{align}
t&=\tau \\
x &= (X - U_0 \tau)/a_x \\
y &= Y/a(t) \\
z &= Z/a(t)
\end{align}
The parameter 
\be
a_x=L_x/L_y^0=L_x/L_z^0
\label{a_x}
\ee
is the initial aspect ratio of the domain ($L_x$, $L_y$, $L_z$ being the size of the domain in the three directions, and the suffix 0 denoting
the initial value). Note that the radial size $L_x$ is a constant, while the other sizes of the domain increase linearly with time.
The parameter $a$ measures this transverse expansion. It is defined as the heliospheric distance $R(t)$ of the barycenter of the plasma domain, normalized by the initial distance $R_0$: 
\be
a=R(t)/R_0=1+\epsilon t = L_y/L_y^0 = L_z/L_z^0
\label{aaa}
\ee
In this equation, $\epsilon=da/dt$ is the \textit{expansion parameter} defined as the initial ratio between the characteristic expansion and 
turnover times in the transverse directions (perpendicular to the radial):
\be
\epsilon=\frac{\tau_{NL}}{\tau_{exp}}=\frac{U_0/R_0}{k_{y}^{00} u_{rms}^0} 
\label{epsi}
\ee
where $k_y^{00}=2\pi/L_y^0=2\pi/L_z^0$ is the initial minimum wavenumber in the transverse directions and $u_{rms}^0$ is the initial root mean square value of the velocity fluctuations. 

The EBM equations finally read, with dissipation terms omitted:
\begin{align}
&\partial_t \rho + \nabla (\rho \vec{u}) = -2  \rho (\epsilon/a)
\label{eqro} \\
&\partial_t P + (\vec{u}.\nabla)P + \gamma P \nabla.\vec{u} = -2  \gamma P (\epsilon/a) 
\label{eqdtp}\\
&\partial_t \vec{u} + \vec{u}.\nabla \vec{u} + \nabla (P+B^2/2)/\rho - \vec{B}.\nabla \vec{B}/\rho = - \vec{\mathbb{U}} (\epsilon/a) 
\label{eqU}\\
&\partial_t \vec{B} + \vec{u}.\nabla \vec{B} - \vec{B}.\nabla \vec{u} + \vec{B} \nabla.\vec{u} = -\vec{\mathbb{B}} (\epsilon/a)
\label{eqB} \\
&P = \rho T
\label{eqP}
\end{align}
where $\mathbb{U}=(0,u_y,u_z)$ and $\mathbb{B}=(2B_x,B_y,B_z)$.
The nabla operator that appears in the previous equations is written in terms of comobile coordinates as 
$\nabla=(1/(a_x)\partial_x,(1/a(t))\partial_y,(1/a(t))\partial_z)$. 
The plasma is transported radially, which implies that, in a local cartesian coordinate system (x,y,z) where x represents the local radial direction, it expands in directions y and z perpendicular to the local radial direction (fig.~\ref{febm}).

Expansion modifies the evolution in two ways: (i) by damping the different fields amplitudes (cf. the rhs terms in eq.~\ref{eqro}-\ref{eqP}); (ii) by damping the gradients perpendicular to the radial, due to the $1/a(t)$ factor in the nabla expression.

Explicit visco-resistive terms (not shown) are added to the previous equations in order to be dissipate the energy that is transported along the spectrum down to the smallest available scales. This energy is given to the internal energy, so that the heating per unit mass reads:
\be
Q_\nu = \mu (\tilde \omega^2+ 4/3 \ (\tilde \nabla\cdot u)^{2}) + \eta \tilde J^2
\label{qnu}
\ee
where $\tilde \omega = \tilde \nabla \times u$  is the vorticity, $\tilde J = \tilde \nabla \times B$ is the current density and 
$\tilde{\nabla}=( \partial_x,\partial_y,\partial_z)$ is the nabla operator defined with respect to the comobile coordinates: this choice of the comobile coordinates for dissipation terms ensures a better control of energy at grid scales.

The dissipation appears as a non-adiabatic source term $\rho Q_\nu$ in the equation for the internal energy, which reads:
\be
\partial_t P + (\vec{u}.\nabla)P + \gamma P \nabla.\vec{u} +2  \gamma P (\epsilon/a) =
\bar \rho \kappa \tilde \Delta T+ (\gamma-1) \bar \rho Q_\nu
\label{eqp}
\ee
where $ \kappa$ is the thermal conductivity.

The corresponding heating is negligible at start, when energy is concentrated in large scale fluctuations only, but becomes substantial, as soon as nonlinear couplings have transferred energy towards scales small enough. The heating (and associated dissipation) can then be called ``turbulent'', 
and, at the same time, gains universality, being in principle independent of the precise value of viscosity and resistivity, and is believed to be a correct prediction for the interplanetary plasma as well, where dissipation is not achieved by visco-resistive terms.

\subsection{Basic physics}
\label{basic}
\subsubsection{Expansion vs turbulence, generalities}
\label{epsaniso}
We summarize here the respective effects of the mean radial wind (also called ``expansion'' in the following) and of turbulence, as discussed in particular in \cite{1993PhRvL..70.2190G,Grappin:1996ey,Dong:2014fi}, \citetalias{VG16} and \citetalias{MGV18}.

The mean radial wind alone is responsible for the following effects: 
(i) eddies become elongated in the two directions perpendicular to the radial directions;
(ii) fluctuations are damped;
(iii) plasma cools.

On the contrary, turbulence alone is responsible of the following:
(i) eddies become elongated in the direction of the mean magnetic field; 
(ii) energy cascades from large to small scales, leading to a turbulent dissipation, which heats the plasma.

The respective importance of expansion and turbulence is measured by the expansion parameter $\epsilon$ which
is the ratio of nonlinear time over expansion time (eq.~\ref{epsi}).
When $\epsilon=0$, the system evolves as in standard MHD. 
When $\epsilon \ll 1$, the largest eddies nonlinear time is much smaller than the transport time, so that turbulence is expected to evolve in the same way as without expansion, that is, expansion simply adds its own decay laws to turbulent dissipation.
Note however that in the limit of vanishing $\epsilon$, the number of nonlinear times to be integrated increases without limit, and we cannot hope to study the turbulent evolution from 0.2 and 1 AU in a finite computational time.

When expansion is strong enough, when, e.g., $\epsilon \simeq 1$, one expects that it will modify the turbulent evolution, as e.g., by delaying shock formation \citep{1993PhRvL..70.2190G}, and/or by forming anisotropic turbulent structures with radial symmetry (\cite{Dong:2014fi}, \citetalias{VG16}). {It is worth mentioning  that expansion might also lead to the formation of magnetic field switch-backs abundantly observed during the first two PSP encounters \citep{Squire_2020}. 
}

\subsubsection{Diagnostic tools and notations}
Spectral anisotropy is basic here, as we will consider it to be a signature of either slow winds or fast winds, according to the findings of 
\cite{2005ApJ...635L.181D}.
Simulations can provide either 3D energy spectra $E^M_{3D}(k_x,k_y,k_z)$ (the suffix M denoting the magnetic fluctuations) or its inverse Fourier transform, the 3D autocorrelation $AC$:
\be
AC(\mathcal{L}_x,\mathcal{L}_y, \mathcal{L}_z)
=\int E_{3D}^M(k_x,k_y,k_z)e^{i(k_x \mathcal{L}_x+k_y \mathcal{L}_y 
+ k_z \mathcal{L}_z)}dk_x dk_y dk_z.
\label{acdef}
\ee

We will also consider the reduced 1D energy spectra in the three x,y,z directions. We thus define:
\be
E_{1D}(k_x)=\int E_{3D}(k_x,k_y,k_z) dk_y dk_z
\label{E1D3D}
\ee
and similarly for the other two spectra $E_{1D}(k_y)$ and $E_{1D}(k_z)$. 
Without special mention, the 1D spectra will show total energy, that is kinetic + magnetic energy per unit mass.
The x direction is that of the radial, with \cha{the x,y} plane containing the mean magnetic field direction. 

In the following, we will always use physical and not comobile coordinates.
Units of wavenumbers will be the initial smallest transverse wavenumber $k_y^{00} = 2\pi/L_y^0$, 
where $L_y^0$ is the transverse size of the initial numerical domain.
Accordingly the units of spatial lags $\mathcal{L}_{x,y,z}$ will be $2\pi/k_y^{00}$.

\subsubsection{Heating}
The critical heating that is necessary to produce the slow cooling of temperature as observed by 
\cite{1995JGR...100...13T} can be obtained as follows.
Imposing a mean temperature profile $\bar T \propto 1/R$ in eq.~\ref{eqp} and taking the spatial average, one finds after some simplifications that the heating rate must be $Q_\nu=Q_c$ where the ``critical heating'' resulting from a turbulent cascade must be:
\be
Q_c = (1/2) \bar T U_0/R
\label{qcrit}
\ee
This formal solution is a slightly modified version of the one considered in \cite{1995JGR...10019839V}.

Now, we may rewrite the critical condition $Q_\nu=Q_c$ in terms of $M$ (turbulent Mach number) and $\epsilon$.
To do this, we use $Q_{K41}$, the phenomenological estimate of the energy transfer rate (equal to $Q_\nu$) in the inertial range:
\be
Q_{K41} = k(u^2 + \delta B^2/\rho)^{3/2} \simeq 3ku^3,
\ee
where k is within the inertial range and $u^2$, $\delta B^2/\rho$ are the kinetic and magnetic energies of fluctuations within the range $[k/\sqrt 2, k \sqrt 2]$, which are assumed to be equal.
\citet{2007JGRA..11207101V} have shown that the K41 expression is related in cold winds to the turbulent dissipation rate as:
\be
Q_{K41} \simeq 10 \bar{Q_\nu}
\label{vasquez}
\ee
So that after simple algebra the critical condition $Q_\nu=Q_c$ (cf eq.~\ref{qcrit}) becomes
\begin{equation}
M^2/\epsilon \simeq 4.4
\label{m2e}
\end{equation}
where we define the turbulent Mach number as
\be
M= u_{rms}/\sqrt{(5/3) T}
\ee

The non-dimensional ratio (eq.~\ref{m2e}) has been used to initialize simulations when studying the turbulent heating of slow/cold winds in \citetalias{MGV18}.
In this paper, we found that eq.~\ref{vasquez} was approximately verified in our direct MHD simulations.
Typically, we used $M=1$, $\epsilon=0.2$.

\section{Initial conditions}
\label{iconditions}

\subsection{GYRO and ISO symmetries, domain aspect ratio}

\begin{table}
\begin{tabular}{ccccccccccc}
Run & Type & N & $\sigma_{c}^0$ & $\epsilon$  &$a_x$ &$\mu_0 [ 10^{-3}]$\\
\hline
G2&GYRO	&512&	$0$		 	&  $0.2$	&5    	& $1.5$\\ 
I2 & ISO		&512&	$0$		 	&  $0.2$	&1	&$3.5$\\ 
I4 & ISO		&512&	$0$			&  $0.4$	&1    	&$2.5$\\ 
I6 & ISO    	&512&	$0$			&  $0.6$	&1    	&$4.6$\\ 
I8 & ISO		&512&	$0$			&  $0.8$	&1    	&$7$ \\  
I4s &ISO   	&512&	$0.95$	 	&  $0.4$	&1	&$4.5$\\ 
I4s5 &ISO 	&512&	$0.95$	 	&  $0.4$	&5	&$1.5$\\ 
I4s5M &ISO	&1024&	$0.95$	 	&  $0.4$	&5	&$1.0$\\  
\end{tabular}
\caption[]
{
Parameters and initial conditions for the simulations. From left to right: name of the run;  type of 3D symmetry of the spectra (ISO for isotropic, GYRO for gyrotropic, i.e. \cha{approximately axisymmetric around the mean magnetic field}, see section \ref{iconditions}); number of grid points $N$ in each direction; the normalized cross helicity, $\sigma_{c}^0$; the expansion parameter, $\epsilon$; the aspect ratio of the numerical domain, $a_x$; the value of the dynamic viscosity, $\mu_0$.
\cha{For all simulations}, the maximum excited wavenumbers are $k^{max}_{y,z}/k_0=4$ for the directions transverse to the radial and $k^{max}_{x}/k_y^{00}=a_{x}^{-1} k^{max}_{y,z}/k_y^{00}$ in the radial direction; the 1D spectral index is $m_0=5/3$, the turbulent Mach number is $M_0=1$, with $u_{rms}=b_{rms}=1$, and the mean magnetic field is slightly inclined in the ecliptic plane, ${\bf B}_0=[2,2/5,0]$. 
}
\label{table1}
\end{table}

\begin{figure}
\begin{center} 
\includegraphics [width=0.4\linewidth]{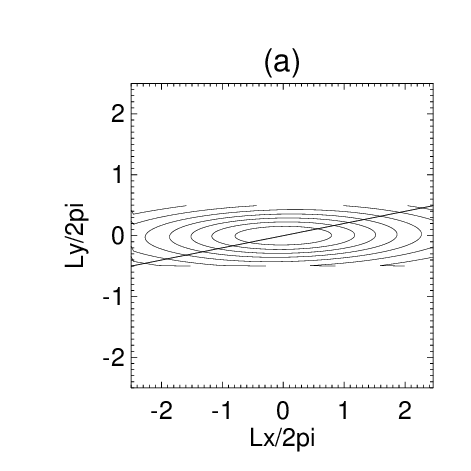}
\includegraphics [width=0.4\linewidth]{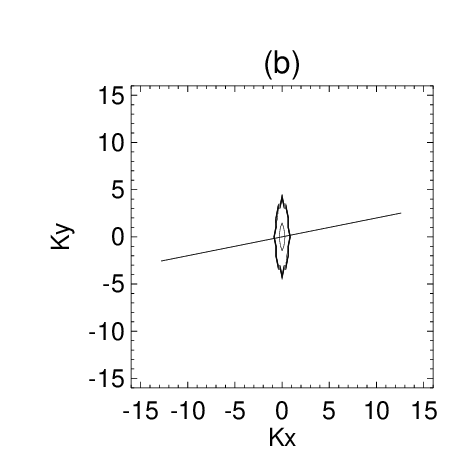}
\includegraphics [width=0.4\linewidth]{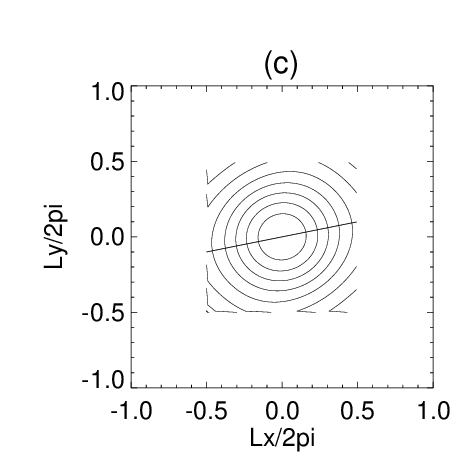}
\includegraphics [width=0.4\linewidth]{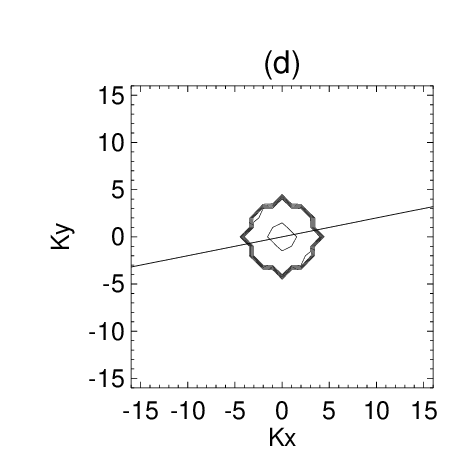}
\caption{
The two kinds of Initial conditions. Top panels: GYRO symmetry; bottom panels: ISO symmetry. 
Left panels: 3D autocorrelations $AC(L_x,L_y,L_z=0)$;
right panels: 3D energy spectra $E_{3D}(k_x,k_y,k_z=0)$.
Straight lines show the mean field direction. Note the limited number of wave vectors excited: 
$1/5 \le k_x \le 4/5$ and $1 \le k_{y,z} \le 4$ in the GYRO case,
$1 \le |k| \le 4$ in the ISO case.
}
\label{fispectra}
\end{center}
\end{figure}
\begin{figure}
\begin{center}
\includegraphics [width=0.6\linewidth]{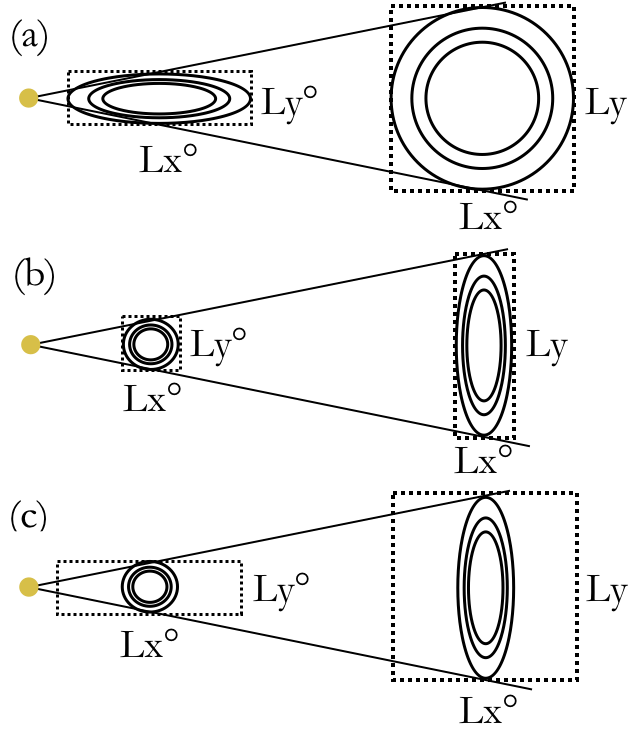} 
\caption{
Initial and final domains of the simulations in the X-Y plane (dotted lines). 
(a) Initial  aspect ratio $a_x=L_x^0/L_y^0=5$, final aspect ratio unity; 
(b) Initial aspect ratio unity, final aspect ratio 1/5;
(c) same as (a) but with different initial autocorrelation.
Inside each domain we represent schematically the autocorrelation corresponding to the initial GYRO and ISO symmetry, (a) and (b) respectively, and the expected symmetry at 1~AU when only \textit{linear} effects are accounted for.
}
\label{figGI}
\end{center}
\end{figure}

We \cha{consider as in \citetalias{VG16} two kinds of initial conditions at 0.2 AU, denoted ISO and Gyro in the following.
The ISO 3D spectrum is isotropic.
The Gyro 3D spectrum is an ellipsoid with minor axis parallel to the radial direction, thus quasi-parallel to the mean field direction, and it corresponds to a quasi-2D geometry\footnote{Choosing a symmetry axis strictly in the mean-field direction does not change the results and only makes initial conditions more complicated. This initial condition is practically equivalent to a quasi-2D geometry.}.
In \citetalias{VG16}, we have shown that, with appropriate choice of parameters, the ISO structure transforms into a \textit{radial-slab} structure.
On the contrary, the Gyro structure is seen to keep its initial ellipsoid structure, with its minor axis following the mean magnetic field as it rotates with increasing heliocentric distance, and with its length following the expected classical critical balance between linear Alfv\'en time and nonlinear time \citetalias{VG16, MGV18}.}

Fig.~\ref{fispectra} shows the GYRO (top) and ISO (bottom) initial conditions as 2D cuts in the plane containing the mean magnetic field: the autocorrelations on the left panels and spectra on the right panels.
In this figure, as well as in all other forthcoming figures showing autocorrelations (and spectra), the spatial unit is the \textit{initial transverse} size of the domain divided by $2\pi$.
Note the limited number of wave vectors excited in the spectra:
$1/5 \le k_x \le 4/5$ and $1 \le k_{y,z} \le 4$ in the GYRO case,
$1 \le |k| \le 4$ in the ISO case.
We refer the reader to our previous work \citepalias{MGV18} for a justification of such a choice.

Simulations have $512$ equally spaced grid points in each direction (apart from one run with $1024$ resolution, \cha{see Table~\ref{table1}), }
initial kinetic and magnetic energies are at equipartition, and velocity fluctuations are purely solenoidal.
The initial energy spectra are generated by exciting kinetic and magnetic fluctuations with random phases, with the 1D energy spectrum in the radial direction $E(k_x)$ {being proportional to $k^{-m_0}$, with $m_0=5/3$.}

The initial plasma domain considered in our simulations has either 
an aspect ratio $a_x=L_x^0/L_y^0=5$, thus evolving to a final aspect ratio unity, or
an aspect ratio unity, thus evolving to a final aspect ratio 1/5 (see fig.~\ref{figGI}).
On the contrary, the only run with GYRO initial conditions (G2) has an initial aspect ratio 5 \cha{(see panel a)}, while ISO initial conditions will have $a_x=1,~5$, as in panels b and c.
Note that local couplings are favoured at end or beginning of the calculation depending on the initial aspect ratio being 5 or 1.

\subsection{Choice of parameters}
Consider first the choice of $M_0$ and $\epsilon$.
It is easily found that, for a fixed distance interval, the integration time increases with the number of large scale nonlinear times during travel, that is, when the expansion parameter $\epsilon$ decreases.
For this reason, it was found that $\epsilon$ values within the interval $[0.2,0.4]$ was convenient to begin with
(cf. \citetalias{MGV18}). 

Then, as recalled in the previous section, to reach substantial heating, 
one should adopt a large enough expansion parameter $\epsilon$ and a large enough ratio $M_0^2/\epsilon$ (eq.~\ref{m2e}), which thus leads to $M_0 \simeq 1$. This assumes - as a starting point - that the efficiency of the cascade will be comparable, whatever the geometry of turbulent patterns.

We will then consider initial normalized cross-helicity, $\sigma_{c}^0=0$ or $\sigma_{c}^0 = 0.95$, close to those found for slow and fast winds near 0.3 AU, at solar minimum \citep{2015ApJ...805...84D}. 

The initial (0.2 AU) $b_{rms}$ and $u_{rms}$ equal 1 and the mean magnetic field is $\bb{B}_0=(2,2/5,0)$, close to aligned with the radial.
Due to the conservation of the magnetic flux during expansion by the EBM equations,
the mean magnetic field rotates as distance/time increases until it reaches an angle of $45^{\circ}$ with respect to the radial at the end of the simulation (R= 1 au). 

Table~\ref{table1} lists the runs described in this paper together with the various parameter values: normalized cross helicity $\sigma_c^0$, turbulent Mach number $M^0$, expansion parameter $\epsilon$, initial aspect ratio of the domain $a_x$, initial cutoff wavenumber in the directions transverse to radial $k_{y,z}^{max}$, 1D spectral slope $m_0$, and viscosity $\mu_0$ (assumed to decay as $1/R$ and equal to the resistivity and the thermal conductivity).
Apart from run G2 with GYRO initial anisotropy, the simulations have initial isotropic spectra (ISO symmetry), increasing values of the expansion parameters (runs I2-I8) or of initial cross helicity (runs I4s, I4s5, I4s5M). The latter series are candidates for the radial-slab, and will allow exploring the effect of the initial aspect ratio (passing from 1 to 5) and of resolution (passing from 512 to 1024).

\section{Results}
\label{sresults}
\label{resAni}
In this section we start from the quasi-2D regime that generates a $1/R$ temperature, as studied in \citetalias{MGV18}.
We then change in turn the initial symmetry, the expansion parameter, and the initial cross helicity in order to 
(i) keep a strong heating able to lead to a temperature profile close to  $1/R$;
(ii) pass from the quasi-2D to the radial-slab geometry.

\subsection{From Gyro to ISO initial conditions}
We start with a GYRO initial symmetry, $M^0=1$, and $\epsilon=0.2$ (run G2)\footnote{Such parameters have been used in \citetalias{MGV18} (run E) - the only difference between runs E and G2 is that in the latter the diffusive parameters are somewhat smaller.} and compare it to run I2, which differs only in the symmetry of the initial spectrum, now completely isotropic.
In \citetalias{VG16}, who studied the case of low initial Mach number $M^0=0.12$), isotropic initial conditions were found to enhance the quasi-linear behavior, thus producing an autocorrelation elongated towards the transverse directions instead of towards the mean field in \citetalias{VG16}.

\begin{figure}[th]
\begin{center}
\includegraphics [width=0.45\linewidth] {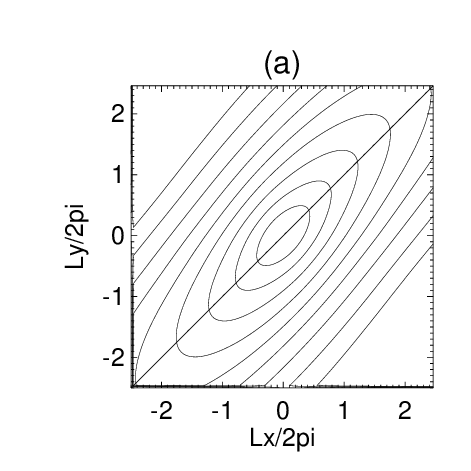} 
\includegraphics [width=0.45\linewidth] {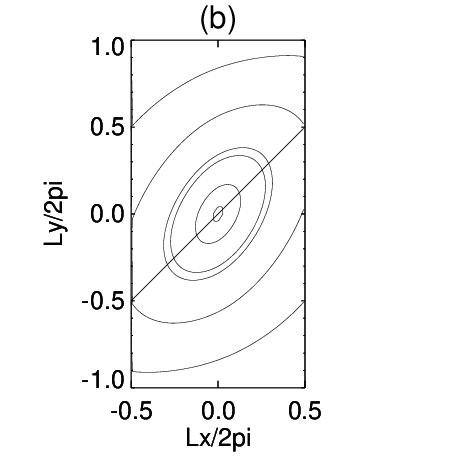} 
\caption{
3D autocorrelation of the magnetic field fluctuations $AC(L_x,L_y,L_z=0)$ at 1 au for simulations with $\epsilon=0.2$ and different initial conditions. \textit{Panel a}. Run G2 (with initial GYRO symmetry). \textit{Panel b}. Run I2 (with initial ISO symmetry).
In each panel, the straight line denotes the mean field direction (here at $45^0$ from the radial direction).
}
\label{figR1}
\end{center}
\end{figure}

In Fig.~\ref{figR1} we compare the magnetic autocorrelations at 1 au in the plane $L_z=0$ for runs G2 and I2.
Panel a shows run G2: at 1 au the autocorrelation's main axis has followed the mean field axis (see straight line) as it rotated away from the initial quasi-radial direction. 
Panel b shows run I2: the ISO initial symmetry leads here at 1 au to a major axis in between the directions of $B_0$ and the $L_y$ axis.
These two results are compatible with those of the low Mach number runs in \citetalias{VG16}.

\begin{figure}[th]
\begin{center}
\includegraphics [width=0.45\linewidth] {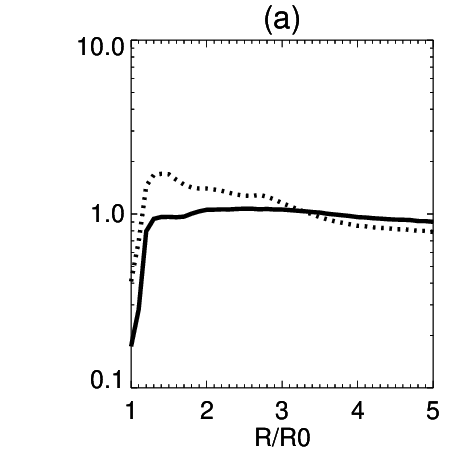}
\includegraphics [width=0.45\linewidth] {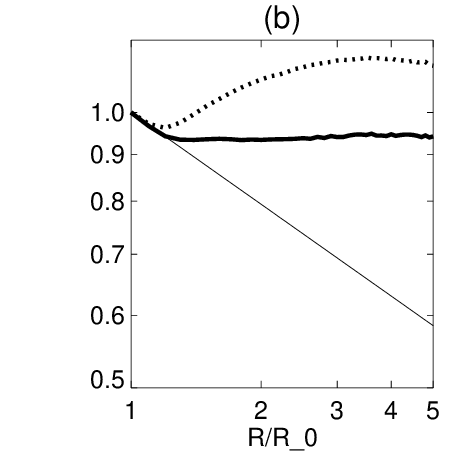}
\caption{
Heating curves for the same runs shown in Fig.~\ref{figR1} with $\epsilon=0.2$ and GYRO symmetry (run G2, thick solid line) and ISO symmetry (run I2, dotted line).
\textit{Panel a}. Heating  $Q_\nu$ vs distance normalized by critical heating $Q_c$.
\textit{Panel b}. Temperature vs distance compensated by $1/R$ decay ($T/T_0 \times R/R_0$), with thin solid line showing the adiabatic solution $1/R^{4/3}$.
}
\label{figR2}
\end{center}
\end{figure}

We then consider in Fig.~\ref{figR2} the heating obtained in both runs, with thick solid lines for run G2 and dotted lines for run I2.
The left panel (a) shows the normalized heating $Q_\nu/Q_c$, and the right panel (b) shows the temperature decay compensated by a $1/R$ decay.
For the GYRO run G2, heating is close to critical everywhere except for a very short phase during which the spectrum forms ($R \le 0.24$~au). 
Correspondingly, the temperature decay is close to $1/R$ except during the short start-up phase for which it follows the adiabatic $1/R^{4/3}$ curve\footnote{These results are similar to those of run E in \citetalias{MGV18}}.
For the ISO run I2, the start-up phase is shorter, with heating being initially well above the critical value, and then becoming somewhat subcritical.
Correspondingly, the temperature decay is close to $1/R$ only in the last interval [0.6, 1] au.

\subsection{Increasing the expansion parameter}
We have seen that imposing isotropic initial conditions turns the symmetry axis of the autocorrelation away from the mean field axis. We now again start with ISO symmetry and increase the expansion parameter $\epsilon$, in the hope to force the final autocorrelation closer to the radial-slab pattern (see \citetalias{VG16}).

\begin{figure}[th]
\begin{center}
\includegraphics [clip, trim= 0 0 1.8cm 0,width=0.3\linewidth] {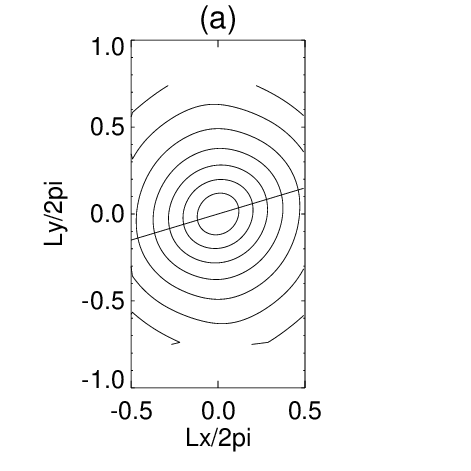}
\includegraphics [clip, trim= 0 0 1.8cm 0,width=0.3\linewidth] {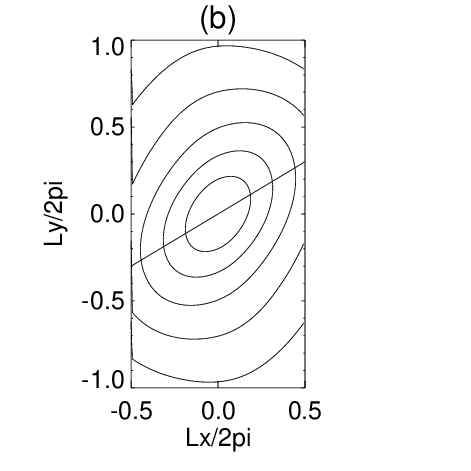}
\includegraphics [clip, trim= 0 0 1.8cm 0,width=0.3\linewidth] {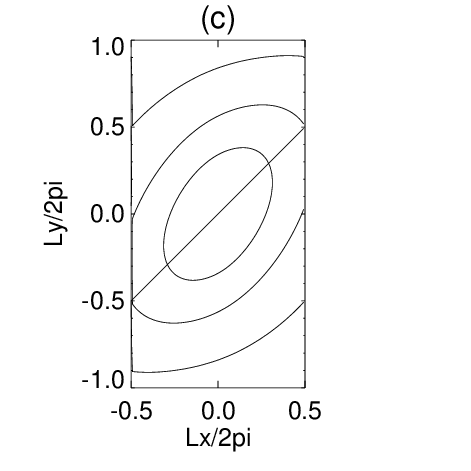}
\includegraphics [clip, trim= 0 0 1.8cm 0,width=0.3\linewidth] {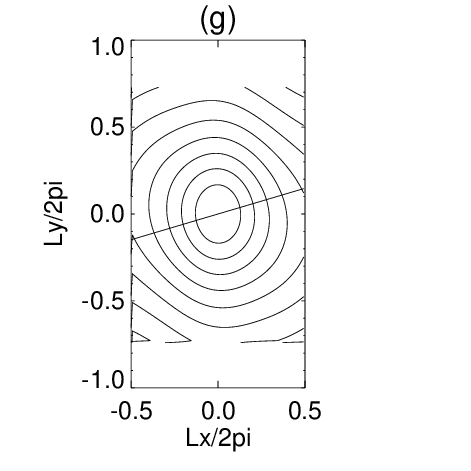}
\includegraphics [clip, trim= 0 0 1.8cm 0,width=0.3\linewidth] {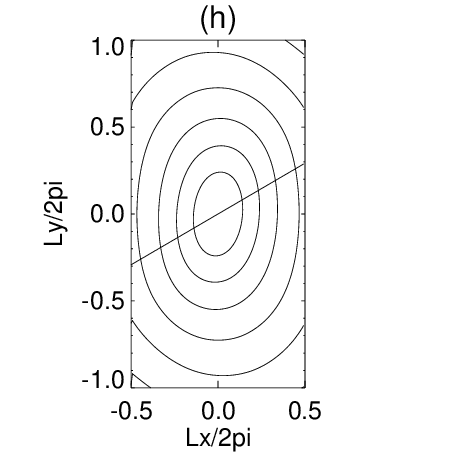}
\includegraphics [clip, trim= 0 0 1.8cm 0,width=0.3\linewidth] {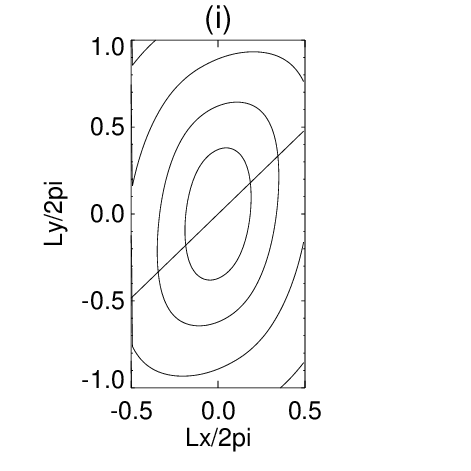}
\caption{
Evolution with distance of the autocorrelation for ISO initial conditions and different expansion parameter. Distance increases from left to right: R=0.3, 0.6 and 1au.
Top panels: run I2 ($\epsilon=0.2$);
Bottom: run I6 ($\epsilon=0.6$).
The straight line in each panel denotes the direction of the mean magnetic field.
}
\label{figR3-4}
\end{center}
\end{figure}

Fig.~\ref{figR3-4} compares the evolution of anisotropy with distance for runs I2 ($\epsilon=0.2$, top panels) and I6 ($\epsilon=0.6$, bottom panels).
While, as already mentioned, run I2 has a major axis in between the $L_y$ and the $B_0$ axis, for run I6 on the contrary, the main axis is now almost parallel to the $L_y$ axis: this is true already at distance $R=0.6$~au (compare panels b and c).

\begin{figure}[th]
\begin{center}
\includegraphics [width=0.45\linewidth] {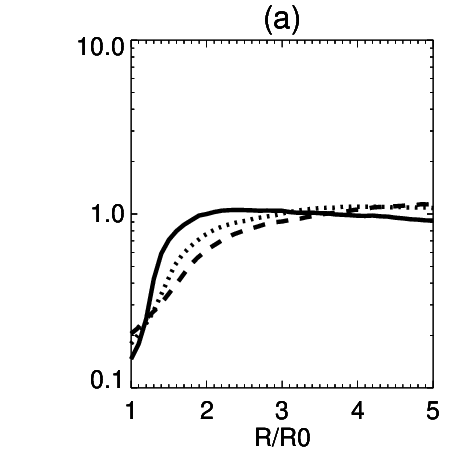}
\includegraphics [width=0.45\linewidth] {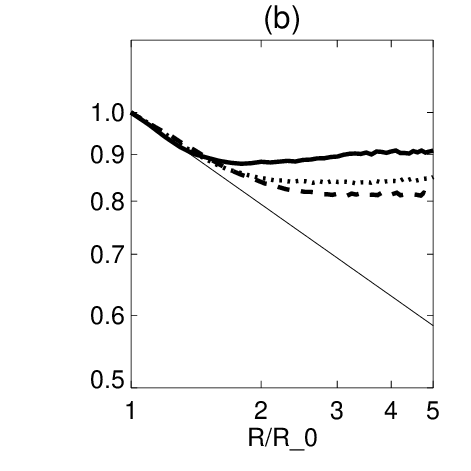}
\caption{
Heating curves for runs shown in Fig.~\ref{figR3-4} (same format of Fig.~\ref{figR2}). ISO symmetry with increasing values of the expansion parameter, $\epsilon=0.4, ~0.6, ~0.8$, for runs I4, I6, I8 in thick solid, dotted, and dashed lines, respectively.}
\label{figR5}
\end{center}
\end{figure}

The heating curves for runs I4, I6, I8 are shown in fig.~\ref{figR5}, in the left panel we plot the heating normalized by the critical heating and in the right panel the temperature is compensated by $1/R$.
Increasing the expansion parameter $\epsilon$ from 0.4 (solid line), to 0.6 (dotted), and 0.8 (dashed), increases 
the delay before a substantial heating occurs, which is expected as the distance traveled during the first nonlinear time becomes larger with larger $\epsilon$.
Note that all three temperature curves show a final stage with approximately an $1/R$ temperature decrease (panel b). 
The largest $1/R$ range is obtained with the smallest expansion, $\epsilon=0.4$.
This is at variance with the case of run I2 considered previously, with $\epsilon=0.2$, which led to early overheating and only in the last distance range an identifiable radial slope for the temperature curve.

\subsection{Non-zero cross helicity}

Since a high cross helicity $\sigma_c$ is frequent, particularly in fast winds, we now compare two ISO runs with respectively $\sigma_c^0$=0 and 0.95 (runs I4 and I4s).
We choose $\epsilon$=0.4, since for this expansion parameter the temperature decays as $1/R$ for $R>0.24$~au and the final anisotropy should be intermediate between that of runs I2 (top figure) and I6 (bottom figure). In other words the autocorrelation's major axis for run I4 should lie in between the mean field axis and the perpendicular to the radial.

\begin{figure}[th]
\begin{center}
\includegraphics [width=0.45\linewidth] {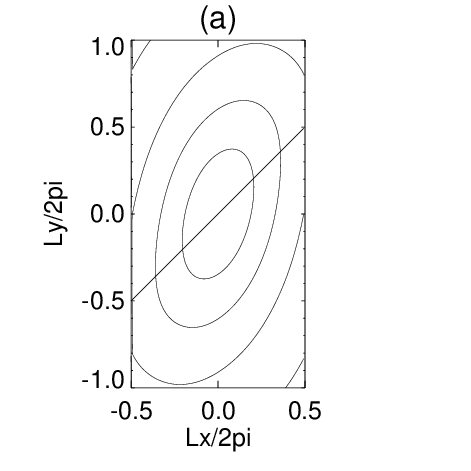}
\includegraphics [width=0.45\linewidth] {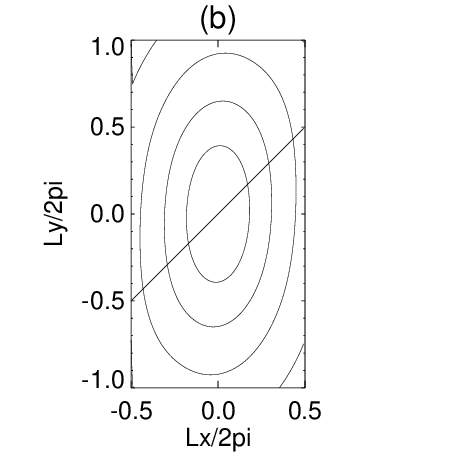}
\caption{Autocorrelation at 1 AU for two runs with $\epsilon=0.4$ and different initial cross-helicity. Panel (a): run I4 with $\sigma_c^0=0$. Panel (b): run I4s with $\sigma_c^0=0.95$.}
\label{figR60}
\end{center}
\end{figure}

\begin{figure}[th]
\begin{center}
\includegraphics [width=0.45\linewidth] {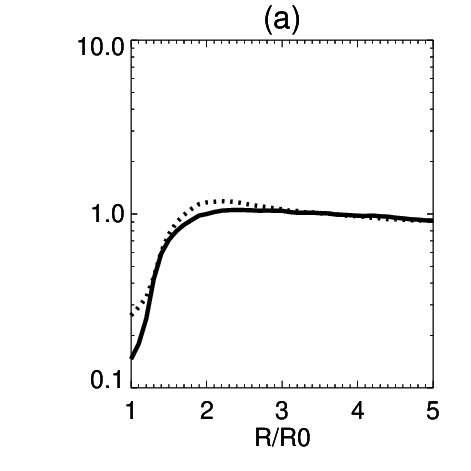}
\includegraphics [width=0.45\linewidth] {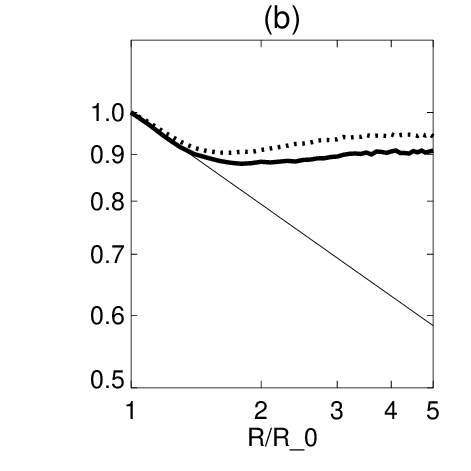}
\caption{
Heating curves for runs shown in Fig.~\ref{figR60} (same format of Fig.~\ref{figR2}). ISO symmetry with vanishing and large cross helicity, run I4 and I4s in solid and dotted lines respectively.
}
\label{figR6}
\end{center}
\end{figure}

Fig.~\ref{figR60} shows the magnetic autocorrelation for runs I4 (a) and I4s (b): a strong cross-helicity clearly moves the major axis closer to the perpendicular to the radial. 
In spite of this difference in anisotropy, we see in fig.~\ref{figR6} that the heating and temperature curves of the two runs don't differ noticeably.

\section{Discussion}
\label{sdiscus}
\subsection{Summary: from quasi-2D to radial-slab}
In order to study the turbulent heating generated by eddies with radial-slab symmetry, we have considered a series of simulations starting from quasi-2D and arriving at quasi perfect radial-slab symmetry, \cha{by keeping the turbulent Mach number equal to one and by only changing the initial configuration at 0.2 AU as follows}:
\begin{itemize}[style=nextline] 
\item changing initial symmetry from GYRO to ISO
\item increasing by a factor two the expansion rate, from 0.2 to 0.4.
\item imposing a strong cross helicity. 
\end{itemize}
\cha{We emphasize that if a pure 2D geometry is chosen as initial conditions, turbulence remains 2D and the plane in which wavevectors are confined does not rotate with the mean field \citep[see][]{2012AIPC.1436..302G}. 
On the contrary, to achieve a rotation of the anisotropy with the mean-field axis,  a quasi-2D initial geometry is necessary, that is, a breath of the spectrum in the mean-field direction (\citealt{2012AIPC.1436..302G}, \citetalias{VG16, MGV18}). 
However, in the following we will see that by adding energy at larger and larger field-parallel wavevectors and by increasing cross helicity, nonlinearities are weakened, but not switched-off, and the symmetry axis unbinds from the magnetic field axis.}

\begin{figure}[th]
\begin{center}
\includegraphics [width=0.32\linewidth] {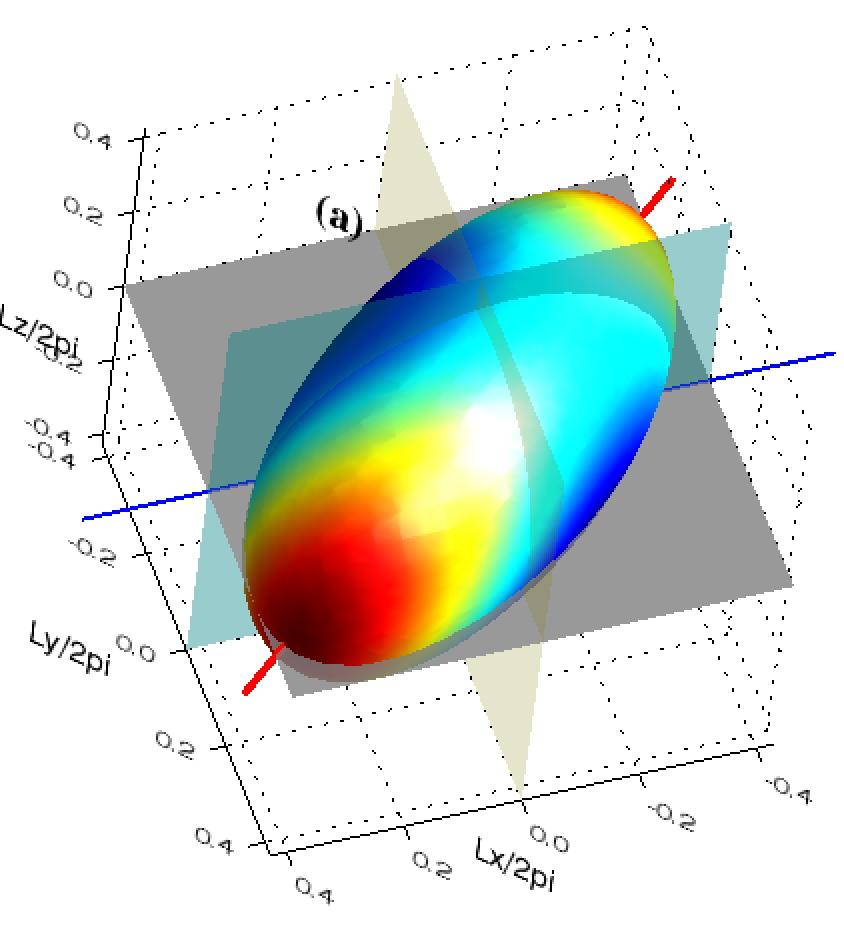}
\includegraphics [width=0.32\linewidth] {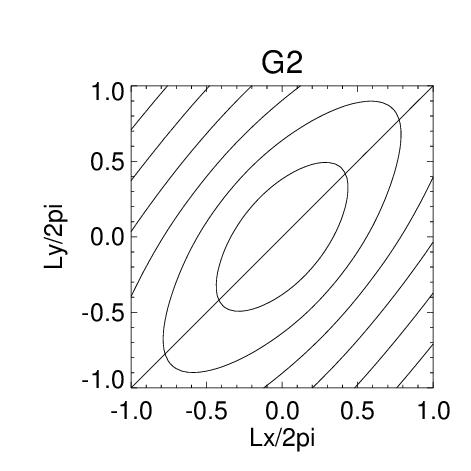}
\includegraphics [clip,trim=0 0 0 1.27cm, width=0.32\linewidth] {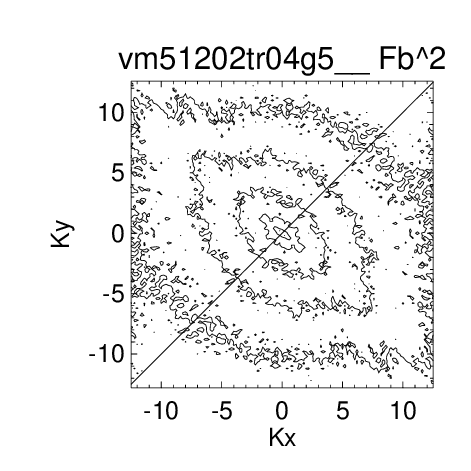} \\
\includegraphics [width=0.32\linewidth] {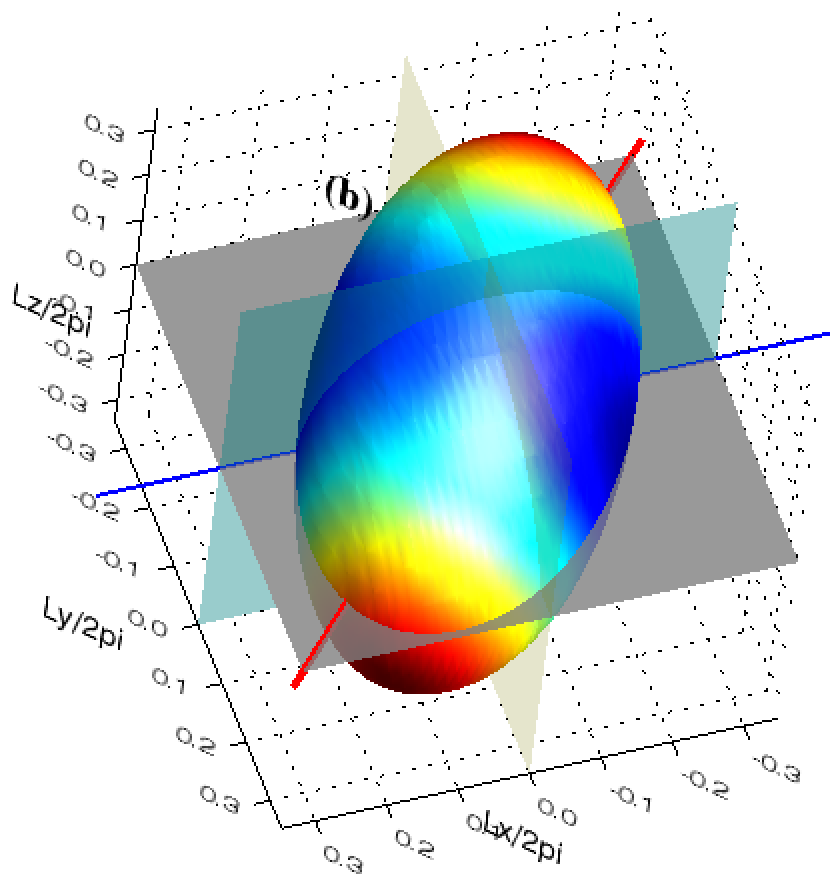}
\includegraphics [width=0.32\linewidth] {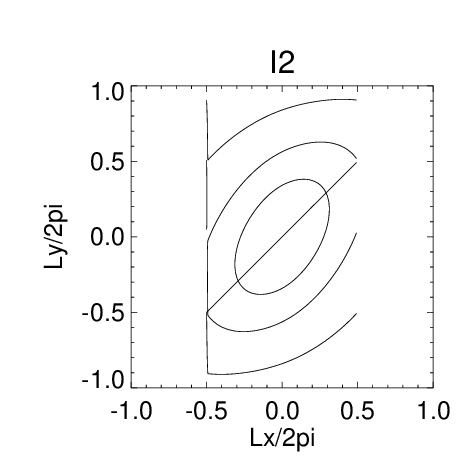}
\includegraphics [clip,trim=0 0 0 1.27cm, width=0.32\linewidth] {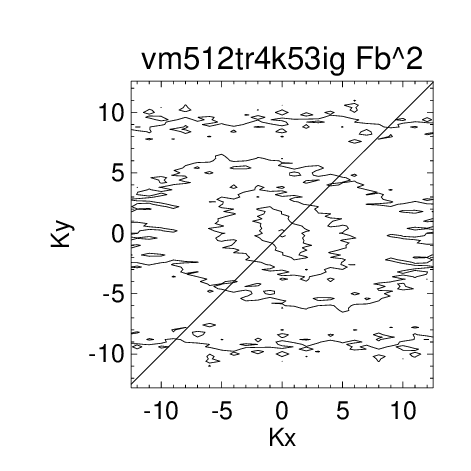} \\
\includegraphics [width=0.32\linewidth] {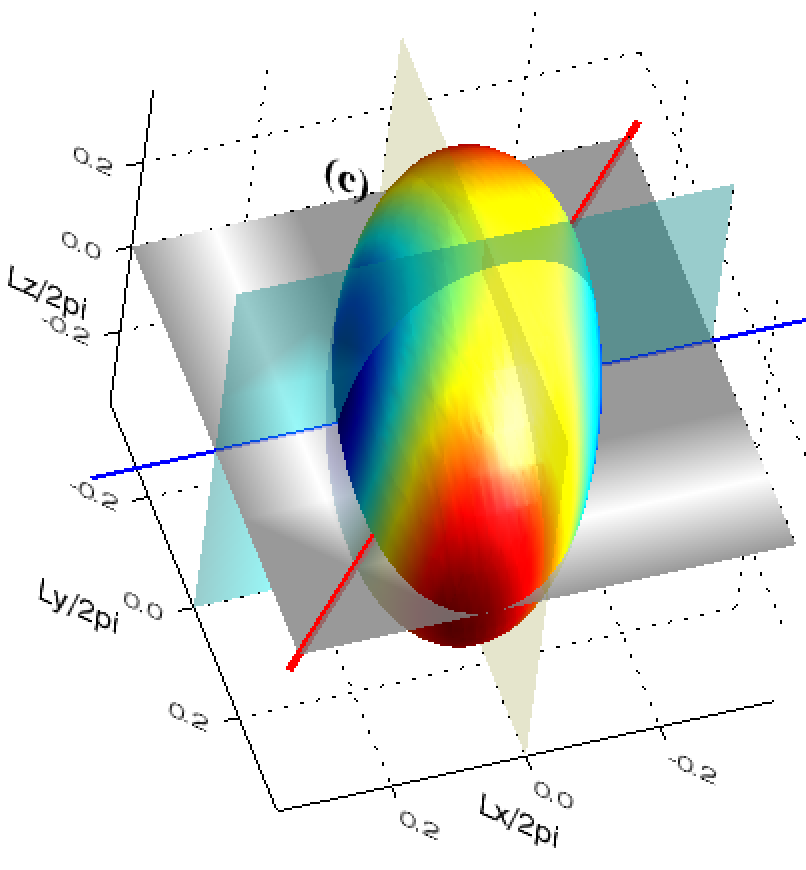}
\includegraphics [width=0.32\linewidth] {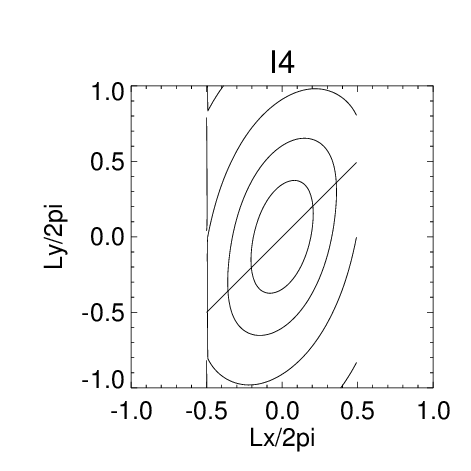}
\includegraphics [clip,trim=0 0 0 1.27cm, width=0.32\linewidth] {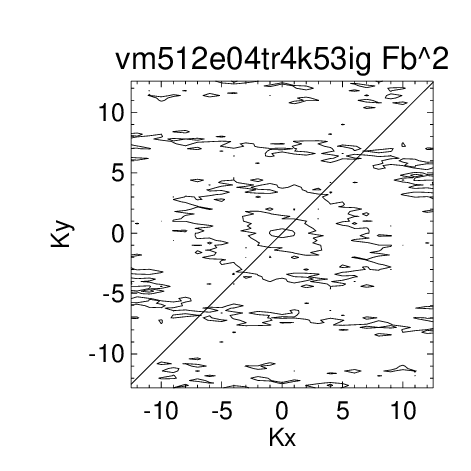} \\
\includegraphics [width=0.32\linewidth] {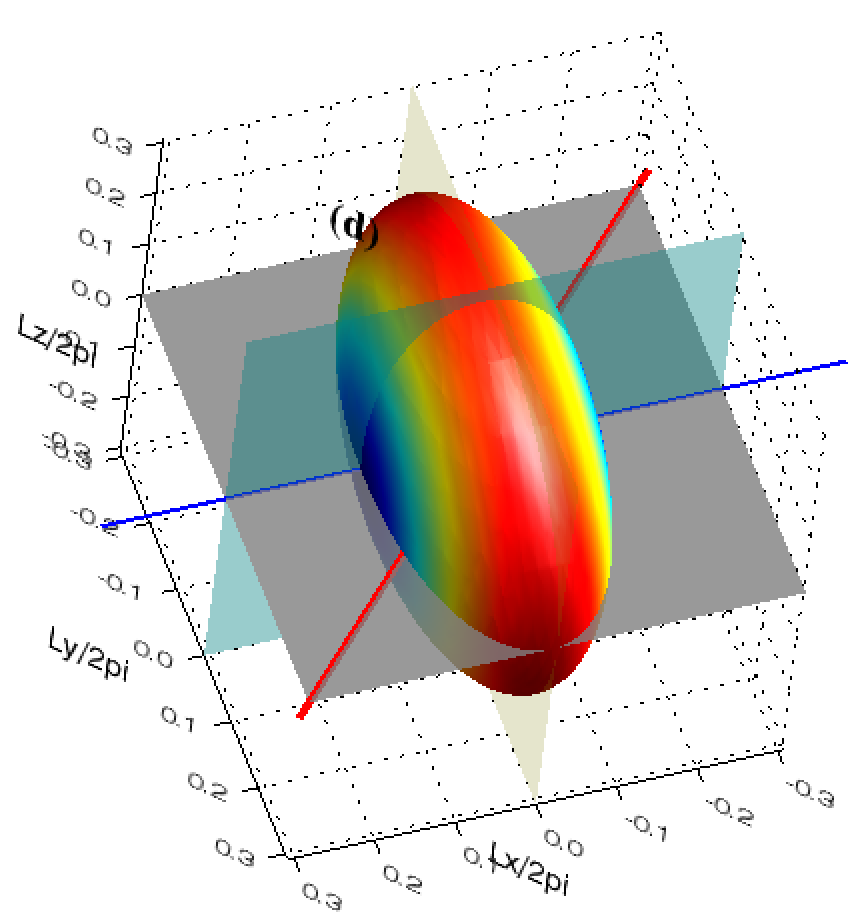}
\includegraphics [width=0.32\linewidth] {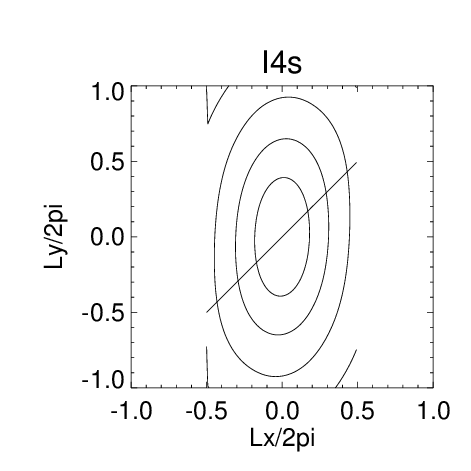}
\includegraphics [clip,trim=0 0 0 1.27cm, width=0.32\linewidth] {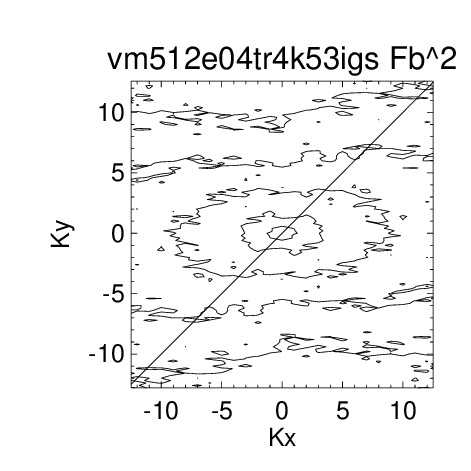}
\caption{Summary of the progressive transition from $B^0-$aligned to radial-slab pattern:
autocorrelation and Fourier spectra.
Left column: given energy isocontour of the 3D magnetic autocorrelation $AC_B$ (with the red line showing the mean magnetic field direction, the blue line showing the radial direction Ox);
middle column: given set of energy isocontours of $AC_B$(Lx,Ly,Lz=0);
right column: given set of energy isocontours of the 3D spectrum, $k_z=0$ plane.
In the last two cases, the straight diagonal line shows the mean field direction.
From top to bottom: runs G2, I2, I4, I4s.
}
\label{figZ1}
\end{center}
\end{figure}

The resulting transition from the quasi-2D to the radial-slab geometry is summarized in fig.~\ref{figZ1}, the runs G2, I2, I4, I4s being shown from top to bottom. 
In each row, we plot from left to right: (a) a given energy level for the 3D magnetic autocorrelation 
$AC_b(L_x, L_y, L_z)$; (b) a set of isocontours of the autocorrelation in the plane $L_z = 0$; (c) a set of isocontours of the spectrum in the plane $k_z = 0$.

In the middle column, the 2D cuts of the autocorrelation show again what we have already seen in the previous section, namely a progressive rotation of the major axis of the autocorrelation from the mean field direction towards the transverse direction. 

In the first column, the 3D iso-surfaces reveal that this is accomplished by a two-fold process: the axis of symmetry changes from the mean-field to the radial direction and at the same time the shape of the autocorrelation gradually changes from prolate to oblate. In other words, for the radial-slab regime (last row) the anisotropy is mainly determined by the transverse stretching due to the expansion of the plasma volume during transport by the wind. 

Autocorrelation best displays the large-scale anisotropy, while Fourier transform best displays medium-scale anisotropy. The two smallest Fourier isocontours in the right column which correspond to the large scales show the same behavior as already seen in the autocorrelation. On the contrary, the large isocontours that correspond to small scales clearly have a tendency to align with the radial axis. 
As we will now, these scales belong to the dissipative range and their alignment with the radial may be ascribed to the absence of nonlinear couplings. Hence, in these results, the radial-slab structure which appears in run I4s seems decoupled from the cascade mechanism.

\subsection{Transverse vs radial spectra}
\begin{figure}
\begin{center}
\includegraphics [width=0.38\linewidth]{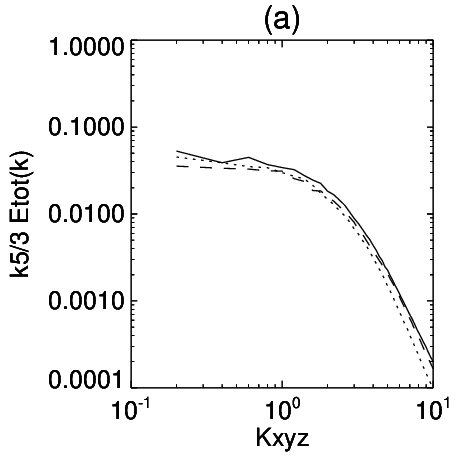}
\includegraphics [width=0.38\linewidth]{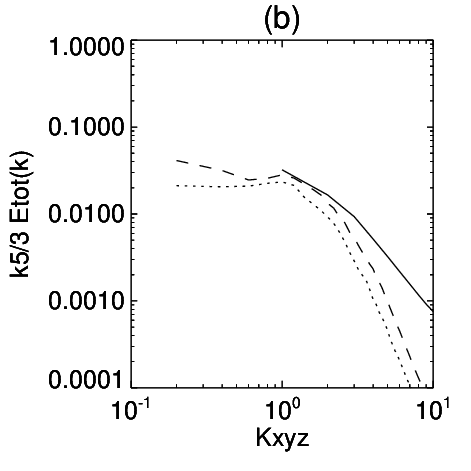}
\includegraphics [width=0.38\linewidth]{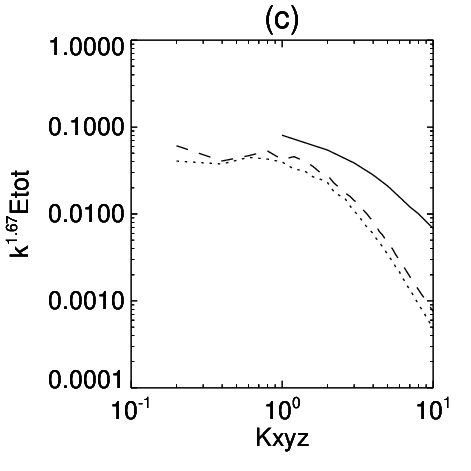}
\includegraphics [width=0.38\linewidth]{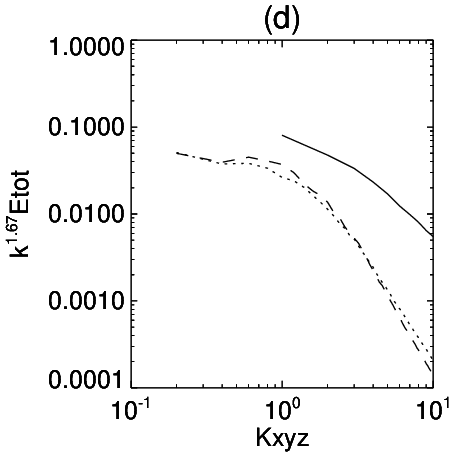}
\caption{1D energy power spectra $E_{tot}(k)$ at 1 AU for the same runs presented in Fig.~\ref{figZ1}: runs G2, I2, I4, I4s in panels (a) to (d) respectively. The 1D spectrum along $k_x$, $k_y$, $k_z$ is plotted with solid, dotted, and dashed lines respectively, and compensated by $k^{-5/3}$.
}
\label{fig1D}
\end{center}
\end{figure}

A better understanding of the radial-slab structure can be gained by looking at 1D spectra. Fig.~\ref{fig1D} shows the total (kinetic+magnetic) 1D energy spectra at 1 au for runs G2, I2, I4, and I4s in panels (a), (b), (c), and (d), respectively. In each panel $E(k_x)$, $E(k_y)$, $E(k_z)$ are plotted with solid, dotted, and dashed lines respectively; all are compensated by $k^{-5/3}$. 
All runs, except perhaps run I2, have transverse spectra $E(k_y)$ and $E(k_z)$ with a scaling index close to -5/3 in the range $0.2 \le k \le 1$. On the contrary, the radial spectrum $E(k_x)$ has only a dissipative tail in the available range $k_x \ge 1$: there is no information at larger scales on the radial spectrum at R=1 AU, for the simple reason that the radial extent $L_x$ of the plasma domain is limited from start to $k_x \ge 1$, and does not change with distance.

In order to reveal the spectral evolution at larger radial scales ($k_x < 1$), we now run again run I4s, but
extending the radial domain by a factor 5, i.e., consider a domain aspect ratio $a_x = 5$ as for the case of the GYRO run. This gives rise to two runs I4s5 and I4s5M, respectively with resolution N=512 and N=1024, the increased resolution of run I4s5M allowing to test the convergence of the results. 
As in previous runs we initialize an isotropic 3D spectrum in the range 
$1 \le k \le 4$ (excluding the modes $k_x < 1$) and use the same random series in runs I4s5 and I4s5M so as to have strictly the same initial conditions. 
\begin{figure}
\begin{center}
\includegraphics [width=0.8\linewidth]{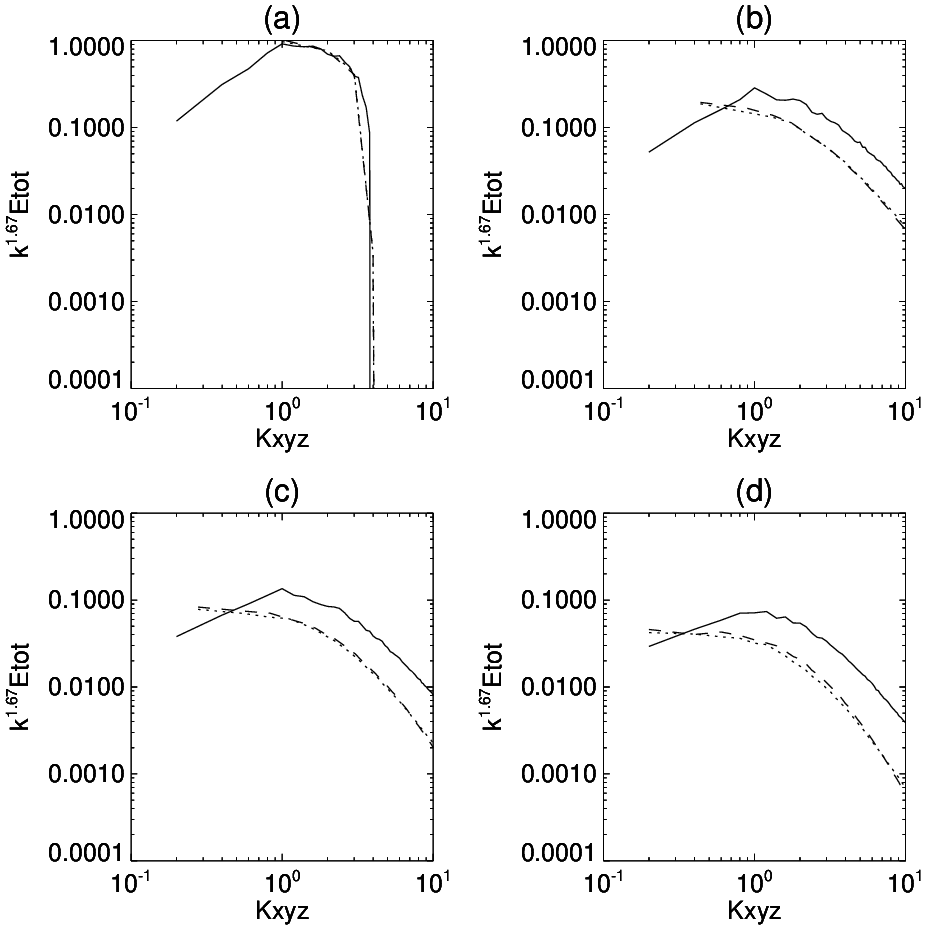}
\caption{Evolution with distance of the three 1D spectra in run I4s5M.
From panel (a) to (d) distance increases as $R=0.2,~0.46,~0.72,$ and 1~au. 1D spectra $E(k_x)$, $E(k_y)$, and $E(k_z)$ are plotted with solid, dotted, and dashed lines, respectively, and compensated by $k^{-5/3}$.
}
\label{fig17M}
\end{center}
\end{figure}

\begin{figure}
\begin{center}
\includegraphics [width=1\linewidth]{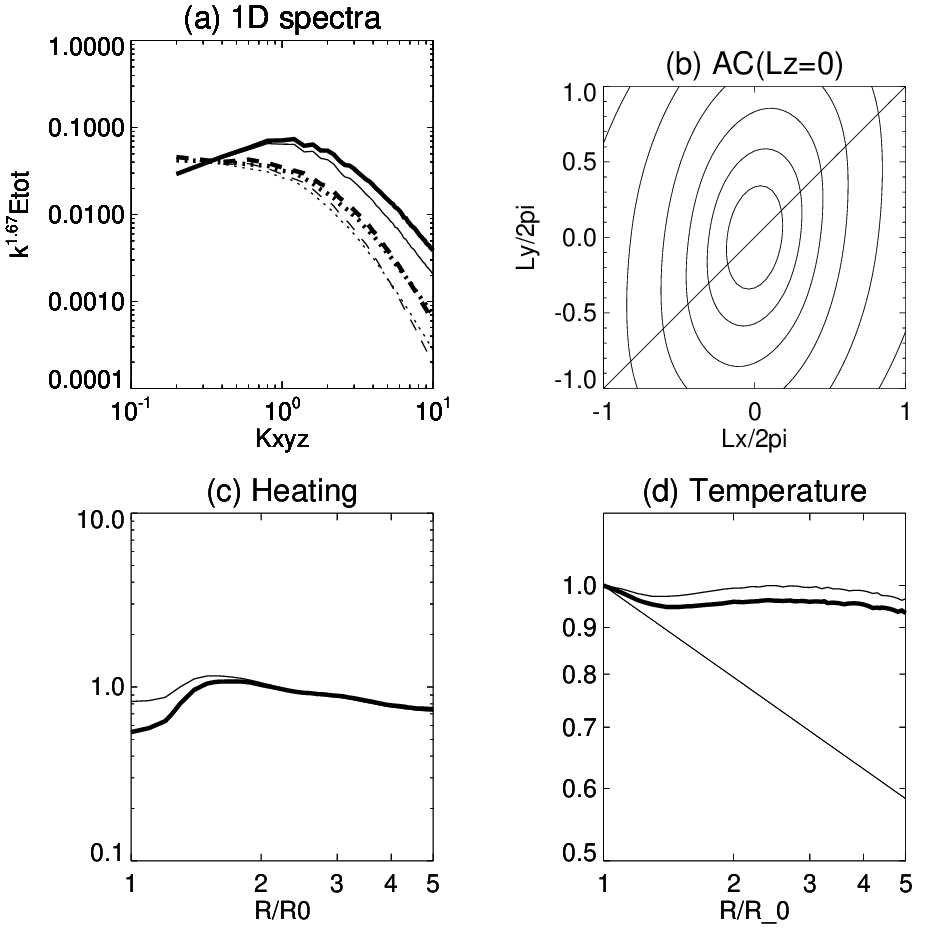}
\caption{Summary plot for runs including large radial scales and ending at 1~au with aspect ratio unity. Runs I4s5M with $N=1024^3$ and I4s5 with $N=512^3$ are plotted in thick and thin lines respectively.
(a) 1D spectra in the three directions at 1 AU, $E(k_x)$, $E(k_y)$, and $E(k_z)$ in solid, dotted, and dashed lines, respectively;
(b) autocorrelation at 1 AU of run I4s5M;
(c) Normalized Heating $Q/Q_c$ vs distance;
(d) Normalized temperature $RT/(R_0T_0)$ vs distance.
}
\label{fig16M}
\end{center}
\end{figure}

First, consider run I4s5M with 1024 resolution. In Fig.~\ref{fig17M} we plot the evolution with distance of 1D spectra in the three directions using the same styles as in Fig.~\ref{fig1D}, with now each panel showing a different distance ($R=0.2,~0.42,~0.76,~1$~AU).
\cha{Note that} the initial spectra (panel a) have the same energy density in the three directions \cha{for $k_{x,y,z} \ge 1$}, which is consistent with
{the isotropy of the 3D spectrum}, however the radial spectrum $E(k_x)$ \cha{is alone to have a} non-vanishing energy in the large-scale range $0.2 \le k_x \le 1$. 
\cha{Recall that this large-scale range exists only in the radial (x) direction. No energy is given to the $k_x$ axis itself (i.e., with $k_y=k_z=0$).
However the situation is different for the lines parallel to the $k_x$ axis (i.e., with $k_y, k_z \ne 0$): these lines cross the shell $1 \le |k| \le 4$, and these lines contribute to the energy spectrum $E(k_x)$ (see eq.~\ref{E1D3D}), which is thus non zero in the range $0.2 \le k_x < 1$.}

During the radial evolution (panels b, c, d) in the range $0.2 \le k_{y,z} \le 1$, the transverse spectra 
$E(k_{y,z})$ progressively flatten, \cha{reaching a scaling index close to $-1.85$ at 1 au}, while the radial spectrum $E(k_x)$ progressively steepens, reaching a \cha{scaling index close to $-1$} at 1 au. Note that the behavior of transverse spectra does not change substantially when changing the initial aspect ratio from $a_x=1$ to $a_x=5$.

An overview of the properties of the two runs with large radial scales is given in Fig.~\ref{fig16M}. 
Runs I4s5 and I4s5M with resolution $512^3$ and $1024^3$ appear in thin and thick lines, respectively. We plot the 1D spectra $E(k_x), E(k_y), E(k_z)$ at 1~au compensated by $k^{-5/3}$ (panel a), the 2D cut of the autocorrelation at 1 AU (only for run I4s5M, panel b), the heating $Q(R)$ normalized by the critical heating 
$Q_c$ (panel c), and finally the temperature profile compensated by $1/R$ (panel d). 
\cha{Panel a shows that the radial spectral index is definitely {-1} for both resolutions, while the transverse spectrum shows some additional flattening when increasing resolution from 512 to 1024.
Panel b shows that the anisotropy is marginally affected by the presence of large radial scales}, with a radial-slab structure that is slightly tilted towards the mean-field direction. Compare with fig.~\ref{figZ1}, central bottom panel (I4s).
Finally, heating is again close to critical and leads to a temperature decrease close to 1/R (panels c and d). 

These properties lead us to conclude that: (i) the heating doesn't vary whether energy is present (runs I4s5M and I4s5) or not (I4s) in large radial scales; (ii) the decoupling between the radial and transverse wave vectors occurs independently of the domain aspect ratio, and (iii) the turbulent heating is due to a turbulent cascade developing in transverse directions, that is, in the plane perpendicular to the radial.

\subsection{Matching Helios and simulation data }
The values of the parameters in our simulations were chosen following two main criteria: (i) an initial expansion parameter $\epsilon$ not too small, in order to limit the computation time, and an initial turbulent Mach number $M_0$ large enough to achieve the required heating (see eq.~\ref{m2e}), in practice $M_0=1$. 
By comparing their values at 1 au with those of Helios 1 mission, we determine now which part of the Solar Wind spectrum these parameters describe.

\begin{figure}
\begin{center}
\includegraphics [width=0.9\linewidth]{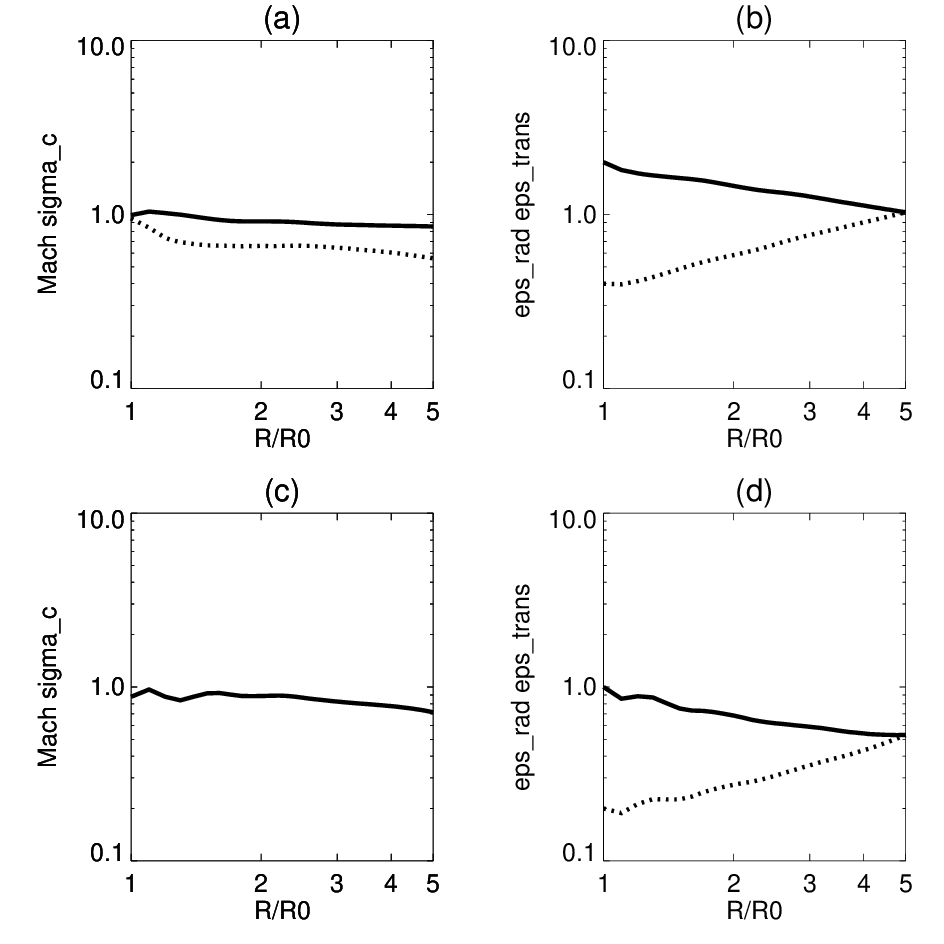}
\caption{
Evolution with distance of several parameters for a radial-slab run (I4s5M, top panels) and quasi-2D run (G2, bottom panels). 
Left panels (a and c): the Mach number $M$ (solid line) and the cross helicity $\sigma_c$ (dotted line) versus distance.
Right panels (b and d): the parameters of radial expansion, $\epsilon_{rad}$ (solid line), and of transverse expansion, $\epsilon_{trans}$ (dotted line) versus distance.
}
\label{fsimul}
\end{center}
\end{figure}

\begin{figure}
\begin{center}
\includegraphics [width=0.8\linewidth]{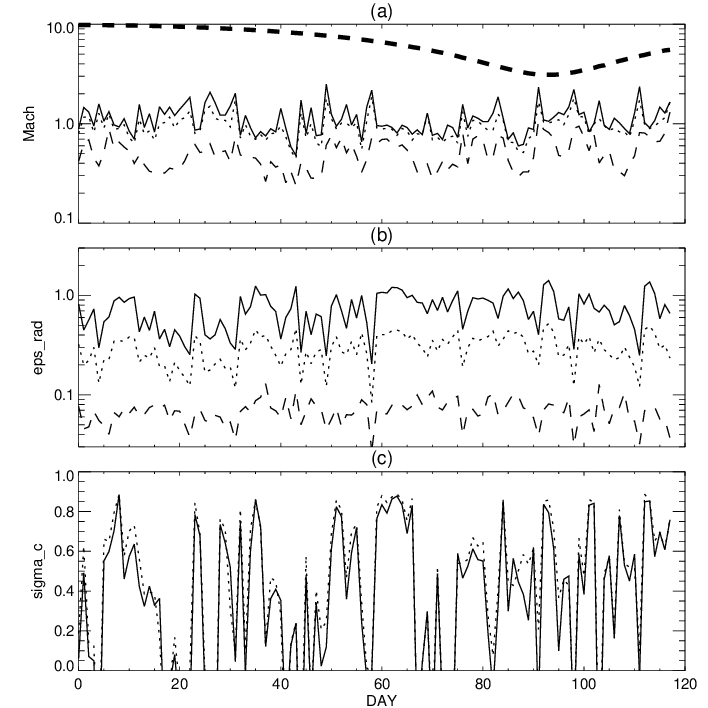}
\caption{
First 118 days of Helios I mission, daily values of (a) turbulent Mach number $M^0$; (b) expansion parameter $\epsilon_{rad}$; (c) cross helicity $\sigma_c$.
RMS quantities are computed on:  $\tau_0$=1$ day$ (solid lines), $\tau_1$=8$h$20 
(dotted lines) and $\tau_2$=1$ hour$ (dashed lines) (a and b only).
Upper panel (a) shows distance $R$ in units of 0.1 au (dashed line).
}
\label{fhelios}
\end{center}
\end{figure}

Fig.~\ref{fsimul} shows the evolution of parameters with distance for the two runs runs I4s5M and G2, representative of the radial-slab and quasi-2D geometries, in the top and bottom panels, respectively.
The turbulent Mach number $M$ is drawn with solid line in panels a and c: it steadily decreases from its initial value $M_0 $~=1 down to 0.85 at 1~au for the radial-slab, and to 0.7 at 1 au in the quasi-2D run.
The cross helicity $\sigma_c$ decreases from $\sigma^0_c = 0.95$ to about 0.6 at 1~au for the radial-slab run (dotted line, panel a), while it remains always close to zero for the quasi-2D run that has a vanishing initial cross helicity (not shown).

In panels b and d, we show the evolution with distance of two relative expansion parameters: the radial and transverse ones. 
The transverse expansion, $\epsilon_{trans}$, is simply defined as the expansion parameter, $\epsilon$ (eq.~\ref{epsi}), in which the nonlinear time is calculated on the transverse box size, taking into account the time/distance variation:
\be
\epsilon_{trans} = \frac{U/R}{k_y^0 u_{rms}}.
\ee
The radial expansion, $\epsilon_{rad}$, is a similar non-dimensional parameter in which the nonlinear timescale is calculated on the largest \textit{radial} scale, and it will be used for comparison with observations, since it allows an evaluation at a fixed Doppler frequency $f$ by applying the Taylor hypothesis:
\be
\epsilon_{rad}=\frac{U/R}{k_x^0 u_{rms}}  = \frac{U/R}{(2\pi f/U) u_{rms}}. 
\label{epsrad}
\ee

As a rule, both $\epsilon_{rad}$ (solid line, panels c and d) and $\epsilon_{trans}$ (dotted line, panels c and d) vary with distance due to the increase of nonlinear time, but $\epsilon_{trans}$ is also affected by the increase of the box size due to expansion, $L_y\propto 1/k_y^0 \propto R$.
Since the domain aspect ratio $L_x/L_y$ reduces from 5 to 1 as distance varies from 0.2 to 1 au, also the ratio 
$\epsilon_{rad}/\epsilon_{trans}$ will vary from 5 to unity. 
The overall evolution is similar in both runs, with $\epsilon_{rad}$ decreasing by about a factor two, so the value at 1~au, $\epsilon_{rad}=1$ and $\epsilon_{rad}=0.5$ for radial-slab and quasi-2D, respectively, is roughly determined by the initial condition. 

We now show in Fig.~\ref{fhelios} the daily evolution of the three parameters $M$ (panel a), 
$\epsilon_{rad}$ (panel b) and $\sigma_c$ (panel c) computed during the first four months of Helios 1 mission in which fast and slow streams are present.
We define averages and fluctuations at three different time scales, 
$\tau_0$ = one day (solid lines), $\tau_1$ =8$h$20 (dotted lines) and $\tau_2$=1~hour (dashed line),
leading to three estimates of $M$, $\epsilon_{rad}$ for these timescales \cha{(for clarity, only the values of $\sigma_c$ at the two largest time scales are given)}. The top panel (a) also shows the heliocentric distance $R$ in units of 0.1 au. 
Note that in single spacecraft data, only the radial scales are measurable by transforming the temporal scale into a spatial scale using the Taylor hypothesis, and so only the parameter $\epsilon_{rad}$ can be obtained.

Table~\ref{table2} allows to compare the parameter values obtained in simulations at 1~au with those found during the first four months of Helios mission. 
For the latter we show the average and standard deviations of $M$ and $\epsilon_{rad}$ values, computed in the range $0.8<R< 1$~au, considering in turn the three timescales $\tau_{0,1,2}$.
For each of the two runs G2 and I4s5M, we indicate in parenthesis the Helios timescales $\tau_{0,1,2}$ for which the Mach and expansion rate obtained in simulations lie within the interval of variation appearing in the three Helios rows.

The comparison of Mach numbers (column denoted by ``M'') shows that the simulation G2 matches Helios scale $\tau_1$ (and marginally $\tau_2$), while simulation I4s5M matches Helios scales $\tau_0$ and $\tau_1$. Using the expansion rate $\epsilon_{rad}$ leads on the contrary to both G2 and I4s5M matching the largest one-day scale $\tau_0$. 
As a final compromise, we choose the following 1 au upper scale: 1 day ($\tau_0$) for the radial-slab (I4s5M) and a half-day, in between $\tau_0$ and $\tau_1$, for the quasi-2D (G2).

\begin{table}
\centering
\begin{tabular}{cccccc}
           &  $\tau$   &$M$               & $\epsilon_{rad}$\\
\hline
Helios & $\tau_0\sim1d$&1.15$\pm$0.4&0.7$\pm$0.3                \\
Helios & $\tau_1\sim8h20$&0.96$\pm$0.3&0.3$\pm$0.1                \\
Helios & $\tau_2\sim1h$&0.5$\pm$0.2  &0.07$\pm$0.02            \\
Quasi-2D (run G2)     &               &   0.7 ($\tau_1$,$\tau_2$)             &	   0.5 ($\tau_0$)                        \\
Radial-slab (run I4s5M) &               &   0.85 ($\tau_0$,$\tau_1$)            &	     1  ($\tau_0$)                
\end{tabular}
\caption[]
{Summary of parameter values measured during the first four months of Helios 1 mission and at the end of simulations G2 and I4s5M.
Top rows denoted by ``Helios'' give the turbulent Mach number $M$ and $\epsilon_{rad}$ with $u_{rms}$ computed on each of the three time scales $\tau_{0,1,2}$, averaged in the interval $0.8 \le r \le 1$~au,  together with their standard deviation.
The two bottom rows give the final values at 1 au of $M$ and $\epsilon_{rad}$ for runs G2 and I4s5M, respectively, together with the time scale(s) ($\tau_{0,1,2}$) for which the simulation results lie within the interval of variation of Helios parameters.
}
\label{table2}
\end{table}

\subsection{Slab versus Radial-slab}

In this work we assumed that the measured anisotropy in the solar wind is well described by a combination of radial-slab and quasi-2D geometry \citep{Saur:1999gy}, leaving aside the more popular combination of slab and quasi-2D geometry \citep[e.g.][]{1990JGR....9520673M,2005ApJ...635L.181D}

As already commented in the introduction, there are several arguments in support of our choice.
First, the analysis leading to the slab geometry relies on the hypothesis of symmetry around the mean magnetic field, which does not necessarily hold in the solar wind. 
Second, by fitting solar wind data at 12h scales to different combination of geometric models, \citet{Saur:1999gy} rejected the slab geometry in favor of the radial-slab geometry, which, in combination with quasi-2D geometry, yielded the most reliable description of fluctuations anisotropy.
Third, there is no numerical support in favor of the slab geometry \citep[but see][]{2017ApJ...835..147Z}, while we showed here and in \citetalias{VG16} that a radial-slab geometry is obtained with isotropic turbulence at 0.2 au and relatively large expansion rate or cross helicity).

In \citetalias{VG16} we also claimed that the radial-slab geometry transforms into a slab geometry when axisymmetry around the mean field is assumed, as in the above-cited observational works that employ a slab geometry. This would represent a definite proof that the underlying geometry is the radial-slab, but it is not the case.
In fact, our claim was based on an erroneous use of 2D FFT. When using 3D FFT the radial-slab geometry transforms into an Isotropic geometry, except in the trivial situation in which the mean field direction is close to the radial direction. This is on average true for the intervals analyzed by \citet{2005ApJ...635L.181D}, the mean-field-angle being $\theta_{BV}\sim30^o$, but since the distribution of angles is broad it cannot be taken as a proof of our claim. 
We cannot thus reconcile the two sets of observations and we defer this to a future work that compares in detail solar wind data and simulations.

\subsection{Conclusion}

In the present work, we have shown that varying the spectral anisotropy, relative expansion rate, and cross helicity at 0.2 au could lead to either 2D, radial-slab or intermediate anisotropy at 1 au, still leading to a temperature decay close to $1/R$.
This numerical proof supports the turbulent origin of the slow temperature decay of solar wind streams \cha{whatever, and in spite of, the observed differences in the spectral anisotropies: (radial-slab or quasi-2D)}.

In the \cha{quasi-2D regime} (run G2), we obtain a scaling close to $k^{-5/3}$ in all directions. 
In the \cha{radial-slab regime} (run I4s5M), we obtain a $1/k$ radial scaling and scalings \cha{close to $k^{-1.85}$} in the two transverse directions. 
The limited resolution doesn't allow to reveal the existence of a possible \textit{radial} $k^{-5/3}$ range at smaller scales and/or at later times.
However, \citet{Bruno:2019hm} showed that cascades starting from scales of one or several days may be found in slow winds, along with cases in which the $1/f$ spectrum shows up. It would be interesting to understand if such differences arise because of different spectral anisotropy, as our simulations indicate.

\cha{Note that either the radial-slab or the quasi-2D regime appears almost immediately, depending on whether the initial symmetry is already isotropic or quasi-2D. In particular the former does not evolve into the latter as in \citet{2019SoPh..294..153R}, possibly due to the absence of shear that would strengthen the cascade.}

Concerning the differences in spectral indices in the radial and transverse directions, it is worth noticing that also the variance anisotropy measured at 1 au was shown to vary with the sampling direction \citep{2016ApJ...832L..16V}: one can expect that also different power-law indices coexist. 
The same combination of -5/3 in the perpendicular direction and -1 in the parallel direction has been found in multishell simulations of the accelerating region \citep{2012ApJ...750L..33V}, suggesting that such initial spectral anisotropy can be maintained at larger distances. Recent measurements taken during the first two Parker Solar Probe (PSP) orbits showed that the 1/ f spectral range exists below 0.2 au \citep{Chen_2020}. 
Observations close to or inside the Alfv\'enic critical point will eventually shed light on its origin. 

Our simulations suggest that the radial spectrum observed at large scales may not inform us about the nature and rate of the cascade.
More precisely, the slow relaxation of the $1/ f$ spectral range towards a steeper scaling observed in the solar wind does not provide a simple criterion to describe a progressive increase of cascade rate and turbulent heating with heliocentric distance. 
Generally, the flattening of the spectra with increasing wind speed, proton temperature, and cross helicity \citep{1991AnGeo...9..416G}, \citep{2013ApJ...770..125C} or decreasing heliocentric distance \citep{Chen_2020} might not be associated with a decrease of cascade rate neither. 

Finally, it would be interesting to use PSP data to extend the temperature profile back to Sun and find until which distance the empirical correlation between turbulent heating and solar wind properties holds, i.e. 
$Q = 0.5U T /R$ \citep{1995JGR...10019839V}. 
At present, observations during the first two PSP orbits roughly confirm its validity. 
In fact, the cascade rate at distances around 0.1 au is about a factor 100 larger than at 1 au and it is correlated with the wind speed \citep{Bandyopadhyay_2020}. 

Our analysis in this study is limited in several ways: (i) only large scales have been considered, due to numerical constraints (studying smaller scales, thus with shorter nonlinear times would take a longer CPU time, the distance interval being fixed); (ii) we limited ourselves to the MHD framework, i.e., with isotropic temperature, and neglecting the difference between proton and electrons; \cha{(iii) we did not include velocity shear or $|B|\approx const$ magnetic fluctuations that are characteristic of the solar wind \citep[see][for this kind of simulations]{2019SoPh..294..153R}; (iv) we are not able to reproduce the observed typical evolution of the $1/f$ range \citep{doi:10.1029/2009JA015120}, nor the shift of the scale separating the $1/f$ and the $f^{-5/3}$ ranges \citep{Bruno_2013_0}. These aspects will be the subject of following works}. In spite of these important limitations, we think that the present work, together with that of \citetalias{MGV18}, provides a significant step forward: it is the first to evaluate in direct simulations the turbulent heating between 0.2 and 1 au, by taking into account at the same time the expansion of the wind and the full nonlinear couplings\cha{, and to prove that a strong heating can be achieved in the inner heliosphere independently of the type of spectral anisotropy}.

\begin{acknowledgements}

The authors would like to acknowledge Thierry Passot and the referee for their  helpful remarks on the manuscript. 
This work was granted access to the HPC resources of CINES and IDRIS under the allocations 2018-A0050407683 and 2019-A0070407683 made by GENCI.
It has been supported by Programme National Soleil-Terre (PNST/INSU/CNRS). This 
research was also supported by OP RDE project No.
CZ.02.2.69/0.0/0.0/16\_027/0008495, International Mobility of Researchers
at Charles University.

\end{acknowledgements}
%

\bibliography{bibvm2}
\bibliographystyle{aasjournal}



\end{document}